\documentclass[aps,twocolumn,a4paper,showkeys,footinbib,longbibliography]{revtex4-1}
\usepackage[utf8]{inputenc}
\usepackage{filecontents}
\usepackage{natbib}
\usepackage{amsmath,amssymb,bm,amsthm}
\usepackage{subfigure}
\usepackage{graphicx}
\usepackage{dcolumn}
\usepackage{bm}
\usepackage[mathlines]{lineno}
\usepackage{booktabs}
\usepackage{color}
\newcounter{one}
\usepackage{url}

\usepackage{textcase}

\usepackage{colortbl}
\usepackage{tabularx}
\usepackage{verbatim}
\usepackage{multirow}

\usepackage[T1]{fontenc} 
\usepackage{lmodern}
\usepackage{bbm}
\usepackage[utf8]{inputenc}
\usepackage{amsfonts}
\usepackage{array}

\usepackage{bbm}
\usepackage{enumitem}
\usepackage{umoline}
\usepackage[usenames,svgnames]{xcolor}
\usepackage{natbib}
\usepackage[hyperindex,breaklinks]{hyperref}
\hypersetup{
     colorlinks=true,       		
     linkcolor=Navy,          	
     citecolor=Navy,            
     filecolor=Navy,      		
     urlcolor=Navy,           	
    runcolor=cyan,
 }
\usepackage{graphicx}
\usepackage{amsfonts}
\usepackage{amssymb}
\usepackage{amsmath}
\usepackage{bbm}
\usepackage{enumitem}

\newcommand{\tr}[0]{ {\rm tr}}

\newcommand{\half}[1]{{ \rm h}}
\newcommand{\Oorderof}{\mathcal{O}}
\newcommand{\orderof}[1]{\Oorderof(#1)} 

\newcommand{\for}[0]{\quad \textrm{for} \quad}
\newcommand{\with}[0]{\quad \textrm{with} \quad}

\newcommand{\dist}{d}

\newcommand{\co}{{\rm c}}

\newcommand{\diam}{{\rm diam}}

\newcommand{\poly}{{\rm poly}}

\def\beq{\begin{equation}}
\def\eeq{\end{equation}}
\def\nbeq{\begin{equation*}}
\def\neeq{\end{equation*}}
\def\<{\langle}
\def\>{\rangle}

\def\tr{{\rm tr}}

\newcommand{\Cl}{\mathcal{C}}
\newcommand{\Gc}{\mathcal{G}}

\newcommand{\Der}{{\mathcal{D}}}

\newcommand{\ed}{X}

\newcommand{\ban}{\mathcal{B}}
\newcommand{\mi}{{\mathcal{I}}}

\newtheorem{theorem}{Theorem}
\newtheorem{lemma}[theorem]{Lemma}
\newtheorem{corol}{Corollary}
 
\newtheorem{definition}{Definition}  
\newtheorem{prop}[theorem]{Proposition} 

\newcommand{\sectionprl}[1]{{\par\it #1.---}}

\usepackage{mathtools}
\def\multiset#1#2{\ensuremath{\left(\kern-.3em\left(\genfrac{}{}{0pt}{}{#1}{#2}\right)\kern-.3em\right)}}

\begin{document}

\title{Clustering of conditional mutual information for quantum Gibbs states above a threshold temperature}

\author{Tomotaka Kuwahara}
\email{tomotaka.kuwahara@riken.jp}
\affiliation{
Mathematical Science Team, RIKEN Center for Advanced Intelligence Project (AIP),1-4-1 Nihonbashi, Chuo-ku, Tokyo 103-0027, Japan
}
\affiliation{interdisciplinary Theoretical \& Mathematical Sciences Program (iTHEMS) RIKEN 2-1, Hirosawa, Wako, Saitama 351-0198, Japan}

\author{Kohtaro Kato}
\email{kokato@caltech.edu}
\affiliation{ Institute for Quantum Information and Matter, California Institute of Technology, Pasadena, CA 91125, USA
}

\author{Fernando G. S. L. Brand\~ao}
\email{fbrandao@caltech.edu}
\affiliation{ Institute for Quantum Information and Matter, California Institute of Technology, Pasadena, CA 91125, USA
}
\affiliation{Amazon Web Services, USA
}

\begin{abstract}

We prove that the quantum Gibbs states of spin systems above a certain threshold temperature are approximate quantum Markov networks, meaning that the conditional mutual information decays rapidly with distance. 
We demonstrate the exponential decay for short-ranged interacting systems and power-law decay for long-ranged interacting systems.  Consequently, we establish the efficiency of quantum Gibbs sampling algorithms, a strong version of the area law, the quasi-locality of effective Hamiltonians on subsystems, a clustering theorem for mutual information, and a polynomial-time algorithm for classical Gibbs state simulations.


%

\end{abstract}
\maketitle

\sectionprl{Introduction} Quantum Gibbs states describe the thermal equilibrium properties of quantum systems. The advent of quantum information science opened up new investigation avenues in the study of Gibbs states, such as the stability of topological quantum memory~\cite{PhysRevLett.107.210501,PhysRevLett.110.090502,PhysRevLett.111.200501,RevModPhys.88.045005}, thermalization in isolated quantum systems~\cite{popescu2006entanglement,Muller2015,brandao2015equivalence,Tasaki2018,kuwahara2019ensemble,kuwahara2019gaussian}, and Hamiltonian complexity~\cite{poulin2011markov,0034-4885-75-2-022001,Aharonov:2013:GCQ:2491533.2491549,gharibian2015quantum}. Efficient methods to prepare quantum Gibbs states in quantum computers have also found proved useful in giving quantum speed-ups for problems such as semidefinite programming~\cite{8104077,8104076,brandao2017quantum} and quantum machine learning~\cite{biamonte2017quantum,PhysRevX.8.021050,PhysRevA.96.062327,song2018geometry,crawford2016reinforcement}.

Quantum Gibbs states also inherit the locality of their parent Hamiltonians, which allows for an efficient classical description in many cases.  One of the simple characterizations is the exponential decay of bipartite correlation functions, which is true in general one-dimension quantum spin lattices~\cite{Araki1969} and in higher dimensions above a threshold temperature~\cite{Gross1979,Park1995,ueltschi2004cluster,PhysRevX.4.031019,frohlich2015some}. 
Another characterization is that at arbitrary finite temperatures, the mutual information between a region and its complement obeys the area law~\cite{PhysRevLett.100.070502}. Quantum Gibbs states also have efficient representations in terms of tensor networks~\cite{PhysRevB.73.085115,PhysRevB.91.045138}.

In classical systems, there are even stronger structural results for Gibbs states. For instance, the Hammersley--Clifford theorem~\cite{HammersleyClifford1971} states that classical  Gibbs states are equivalent to a class of probability distributions called {\it Markov networks}. They satisfy the Markov property; that is, a site is independent from all others conditioned on its neighbors. Therefore, for classical Gibbs states, all the correlations between two separated vertices are induced by the intermediate vertices connecting them.

Although the notion of conditional probability distribution is missing in quantum systems, we can still generalize 
Markov networks to quantum systems using the (quantum) conditional mutual information: 
\begin{align}
&\mi_\rho(A:C|B)  \notag \\
&:= S(\rho^{AB}) +S(\rho^{BC})  -S(\rho^{ABC})  -S(\rho^{B}) ,  \label{cond_info}
\end{align}
where $\rho^{AB}$ is the reduced density matrix in the subsystem $(AB=A\cup B)$ and $S(\rho^{AB})$ is the von Neumann entropy, namely, 
$S(\rho^{AB}):= - \tr ( \rho^{AB} \log \rho^{AB})$. 
In classical systems, the conditional mutual information becomes zero if and only if the state is conditionally independent. 
In quantum cases, the conditional mutual information is related to the approximate recoverability~\cite{Fawzi2015}; hence, it is widely used as the measure of conditional independence in quantum systems.

\begin{figure}[]
\centering
{
\includegraphics[clip, scale=0.25]{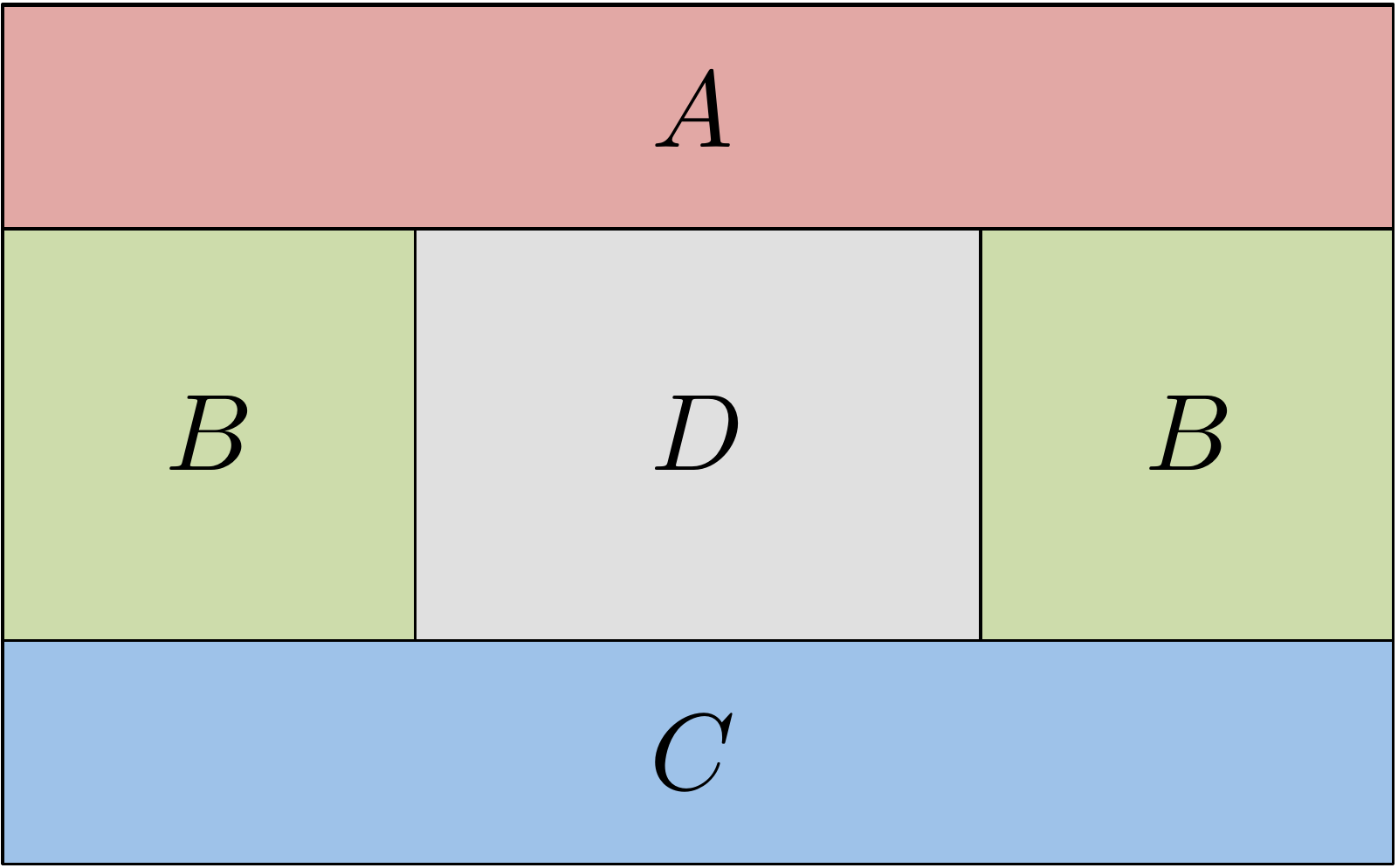}
}
\caption{(color online) Decomposition of the total system into $A$, $B$, $C$, and $D$. 
It is possible that in a quantum state, there is no correlation between $A$ and $C$ when considering only the subsystems $A$ and $C$; however, there is a strong correlation when looking at them via the subsystem $B$.
This kind of correlation between $A$ and $B$ related to $C$ is measured by conditional mutual information~\eqref{cond_info}.
Physically, conditional mutual information characterizes tripartite correlations between $A$, $B$, and $C$. 
A representative example is the topological entanglement entropy~\cite{PhysRevLett.96.110404, PhysRevLett.96.110405}, which is a special form of the 
conditional mutual information.
}
\label{fig:A_B_C_D}
\end{figure}

The quantum version of the Hammersley--Clifford theorem has been established for the case where the Hamiltonian is short-range and commuting~\cite{brown2012quantum,Jouneghani2014}: any quantum Gibbs state of such Hamiltonian on a triangle-free graph is a Markov network and vise versa. 
More recently, it has been shown that the Hammersley--Clifford theorem {\it approximately} holds in one-dimensional lattice~\cite{kato2016quantum}, in the sense that the conditional mutual information of any Gibbs state decays subexponentially with respect to distance.

In the present work, we will establish the approximate Markov property for quantum Gibbs states in spin systems interacting on generic graphs.
In our study, we consider not only short-range interactions but also long-range (i.e., power-law decaying) interactions on graphs. 
We prove that above a certain threshold temperature, the conditional mutual information decays exponentially (polynomially) for short-ranged (long-ranged) models. 
Our result will strengthen the 1D result obtained in Ref.~\cite{kato2016quantum}, the area law for mutual information~\cite{PhysRevLett.100.070502}, and the standard clustering theorem ~\cite{PhysRevX.4.031019,frohlich2015some}. 
Moreover, our result immediately implies a quasi-polynomial-time quantum Gibbs sampling algorithm by following the discussion in Ref.~\cite{Brandao2019}. 
Finally, for computing thermodynamic quantities (e.g., the partition function), we develop a polynomial-time classical algorithm for the first time.




%
\sectionprl{Setup}
We consider a quantum system with $n$ spins, where each spin has a $d$-dimensional Hilbert space. 
We assume that the spins sit on the vertices of a graph $G=(V,E)$ where $V$ is the total spin set ($|V|=n$).
For arbitrary subsystems $A, B \subset V$, we define $\dist_{A,B}$ as the shortest path length on the graph that connects $A$ and $B$. If $A\cap B \neq \emptyset$, $\dist_{A,B}=0$.
We define the surface region of an arbitrary subsystem $L \subseteq V$ as $\partial L_l \subseteq V$ ($l \in \mathbb{N}$): 
\begin{align}
\partial L_l :=\{v\in L| \dist_{v,L^\co}\le l\} , \label{surface_region_of_L}
\end{align}
where $L^\co$ is the complementary set of $L$ (i.e., $L\cup L^\co=V$).

We define the system Hamiltonian $H$ as
\begin{align}
H= \sum_{|\ed |\le k} h_\ed  ,
\label{eq:ham_graph}
\end{align}
where each interaction term $\{h_\ed\}$ acts on the spins in $\ed \subset V$.
The Hamiltonian [Eq.~\eqref{eq:ham_graph}] describes generic $k$-body interacting systems. 
We characterize the locality of the interactions as follows: 
\begin{align}
 \sum_{\substack{\ed | \ed \ni v \\ \diam (\ed) \ge R}} \| h_\ed \| \le f(R) \quad {\rm with} \quad f(1) \le g
 \label{def:interaction_length}
\end{align}
for $\forall v\in V$, where $\|\cdot\|$ is the operator norm and $\diam (\ed)$ is the diameter of $\ed$, namely, ${\displaystyle \diam(\ed):=\max_{\{v_1,v_2\} \in \ed}\dist_{v_1,v_2}}$ for $\ed \subset V$. 
The parameter $g$ corresponds to one-spin energy since $\sum_{\ed | \ed \ni v } \| h_\ed \| \le f(1) \le g$. By taking the energy unit appropriately, we set $g=1$.
For example, if $k=2$ and $f(R)=0$ for $R\ge2$, the Hamiltonian is described by bipartite nearest-neighbor interactions as $H= \sum_{\{i,j\} \in E} h_{i,j}$.
We consider the Gibbs state for the Hamiltonian $H$ at an inverse temperature $\beta$ as follows:
\begin{align}
\rho:= \frac{1}{Z}e^{-\beta H}, \quad Z:= \tr (e^{-\beta H}). \notag
\end{align}

Our purpose is to discuss the Markov property of Gibbs states.
Let $V_0\subseteq V$ be an arbitrary subsystem. Consider a tripartite partitioning of $V_0$ as $V_0=ABC$, where we denote $A\cup B$ by $AB$ for simplicity.
We notice that the subsystems $\{A,B,C\}$ are \textit{not} necessarily concatenated on the graph (see Fig.~\ref{fig:A_B_C_D}).
If any two nonadjacent subsystems $A$ and $C$ are conditionally independent of the other subsystem $B$ ($=V_0\setminus AC$), we say that $\rho^{V_0}$ is the \textit{quantum Markov network} on $V_0$.
Mathematically, this implies $\mi_\rho(A:C|B)=0$ for $\dist_{A,C}>0$~\cite{brown2012quantum,Hayden2004}, where $\mi_\rho(A:C|B)$  is defined in Eq.~\eqref{cond_info}.
It is noteworthy that the Markov property of $\rho^{V_0}$ strongly depends on the selection of the subsystem $V_0\subseteq V$.
To prove this point, let us consider a one-dimensional graph. Then, the GHZ state is a Markov network for $\forall V_0 \subset V$, but not globally, namely, $\mi_\rho(A:C|B)=1$ for $ABC=V$.
In contrast, the cluster state~\cite{PhysRevLett.86.5188} is globally a Markov network, but not for particular selections of $V_0$~\cite{PhysRevLett.122.140506,PhysRevB.94.075151} (see also~\cite{Foot2}). 
Based on the example of the cluster state, which has a finite correlation length and is described by the matrix product state with bond dimension 2~\cite{Perez-Garcia:2007:MPS:2011832.2011833}, 
we cannot ensure the Markov property only using the clustering theorem and matrix product (or tensor network) representation of the quantum Gibbs state.


The Markov property has a clear operational meaning in terms of a recovery map as follows:
If $\rho^{V_0}$ is a Markov network, we can always find a quantum channel $\tau_{B\to BC}$ referred to as the Petz recovery map~\cite{Petz1986,Hayden2004}, which recovers $\rho^{ABC}$ from $\rho^{AB}$ ($V_0=ABC$): 
 \begin{align}
\tau_{B\to BC} (\rho^{AB}) =\rho^{ABC}. \notag
 \end{align}
The above local reconstruction is not possible for generic quantum states. 
Note that any quantum Markov network on a tree graph can be constructed from a sequence of $n$ local quantum channels.

In realistic situations, we frequently encounter cases where the density matrix is not given by the exact Markov network but by an approximate Markov network, that is, 
the conditional mutual information $\mi_\rho(A:C|B)$ approaches zero as the distance $d_{A,C}$ increases. 
In the case where $\mi_\rho(A:C|B)=\epsilon$, the celebrated Fawzi--Renner theorem~\cite{Fawzi2015} (see also \cite{
PhysRevLett.115.050501,
doi101098rspa20150338,
DBLPjournalsqicBertaLW15,
berta2015renyi,
doi101098rspa20150623,
7404264,
Sutter2018,
Junge2018})
ensures the existence of the recovery map such that 
 \begin{align}
\left\| \tau_{B\to BC} (\rho^{AB}) - \rho^{ABC} \right\|_1^2 \le \epsilon \log 2 ,\label{ineq:Fawzi2015}
 \end{align}
where $\|\cdot\|_1$ is the trace norm (i.e., $\|O\|_1:=\tr(\sqrt{O^\dagger O})$ for an operator $O$).
Here, the form of $\tau_{B\to BC}$ is given by the rotated Petz map~\cite{Junge2018}.
Based on this theorem, we can still relate the approximate Markov property to the local reconstruction of the state.

The main purpose of this study was to characterize the decay rate of the conditional mutual information $\mi_\rho(A:C|B)$ with respect to the distance $d_{A,C}$ on the graph.
To explain the physics of the theorems, we have provided the proofs of our main theorems in the supplementary material~\cite{Supplement_Quantum_Markov}.

\sectionprl{Main result}
We proved the exponential decay of the conditional mutual information above a temperature threshold:
\begin{theorem} \label{main_theorem_QAMC_mutual_information}
Let us consider finite-range interaction up to a finite length $r$; that is, we consider a function $f(R)$ in Ineq.~\eqref{def:interaction_length} such that $f(R)=0$ for $R>r \in \mathbb{N}$.
Then, the condition
\begin{align}
\beta < \beta_c := \frac{1}{8e^3 k}  \label{def:beta_c}
\end{align}
implies that the Gibbs state $\rho$ is an approximate Markov network on an arbitrary subset $V_0\subseteq V$ in the sense that
\begin{align}
\mi_\rho(A:C|B)  \le  e\min(|\partial A_r|, |\partial C_r|)   \frac{(\beta/\beta_c)^{\dist_{A,C}/r}}{1-\beta/\beta_c}  ,\label{eq:quantum_approximate_markov_chain}
\end{align}
where $V_0=ABC$ and the subset $\partial A_r$ ($\partial C_r$) is defined by Eq.~\eqref{surface_region_of_L} with $l=r$ and $L=A$ $(L=C)$.
\end{theorem}

We notice that if we select $B$ as an empty set (i.e., $B=\emptyset$), the conditional mutual information reduces to bipartite mutual information:
\begin{align}
\mi_\rho(A:C|\emptyset) =\mi_\rho(A:C), \notag 
\end{align}
where $\mi_\rho(A:C):=S(\rho^{A}) +S(\rho^{C})  -S(\rho^{AC})$.
Therefore, inequality~\eqref{eq:quantum_approximate_markov_chain} also implies the exponential decay of the mutual information between two separated subsystems. 
It is an improved version of the standard clustering theorem for the bipartite operator correlation ${\rm Cor}_\rho (O_A,O_B):= \tr (\rho O_AO_B) - \tr (\rho O_A) \tr(\rho O_B)$, where $O_A$ and $O_B$ are arbitrary operators with unit norm (i.e., $\|O_A\|=\|O_B\|=1$) supported on subsystems $A$ and $B$, respectively.
From the relation $[{\rm Cor}_\rho (O_A,O_B)]^2 \le  2 \mi_\rho(A:B)$~\cite{PhysRevLett.100.070502}, the clustering theorem can be derived from the exponential decay of the mutual information.
Moreover, it is well known~\cite{Hayden2004hiding,PhysRevB.76.035114} in the context of data hiding that even if the operator correlation is arbitrarily small in a quantum state, the state may still be highly correlated
in terms of the mutual information~\cite{PhysRevLett.100.070502}.

An important implication of this theorem is related to the quantum sampling of Gibbs states.
Based on the Fawzi--Renner theorem~\eqref{ineq:Fawzi2015}, an approximate Markov network can be efficiently reconstructed from its reduced density matrix using a quantum computer.
According to Ref.~\cite{Brandao2019}, the clustering and Markov properties ensure an efficient preparation of quantum Gibbs states on finite-dimensional lattices. 
By combining our theorem~\ref{main_theorem_QAMC_mutual_information} with Theorem 5 in Ref.~\cite{Brandao2019}, we obtain the following statement.

Let us consider the case where the graph $G$ is given by a $D$-dimensional lattice, where $D$ is the spatial dimension. Then, under the assumption of Theorem~\ref{main_theorem_QAMC_mutual_information}, 
there exists a completely positive trace-preserving (CPTP) map of $\mathbb{F}= \mathbb{F}_{D+1} \cdots  \mathbb{F}_{2} \mathbb{F}_{1}$ such that
\begin{align}
\|\mathbb{F}(\psi) -\rho \|_1 = 1/\poly(n) ,\notag 
\end{align}
where $\psi$ is an arbitrary quantum state and each of $\{\mathbb{F}_{s}\}_{s=1}^{D+1}$ is given by a direct product of quasilocal CPTP maps that act on $\orderof{\log^D n}$ spins. 

The number of the elementary gates for the each quasilocal channel $\{\mathbb{F}_{s}\}_{s=1}^{D+1}$ is on the order $\exp[\orderof{\log^D n}]=n^{\orderof{\log^{D-1} n}}$~\cite{PhysRevA.52.3457, PhysRevLett.92.177902}. This also provides the computational time of Gibbs sampling by the quantum computer.
This algorithm requires only quasi-polynomial computational time, and it is considerably better than a few existing algorithms~\cite{PhysRevLett.103.220502,PhysRevLett.105.170405}, which require at least subexponential computational time. 
Our algorithm still performs slightly worse than the algorithms proposed in Refs.~\cite{Kastoryano2016} and \cite{PhysRevLett.116.080503}, which require polynomial computational time.
However, our method has advantages in the following senses: the method in \cite{Kastoryano2016} is applicable only to commuting Hamiltonians and the method in \cite{PhysRevLett.116.080503} requires twice the number of qubits (i.e., $2n$ qubits) for implementation.

\begin{figure}[tt]
\centering
{
\includegraphics[clip, scale=0.3]{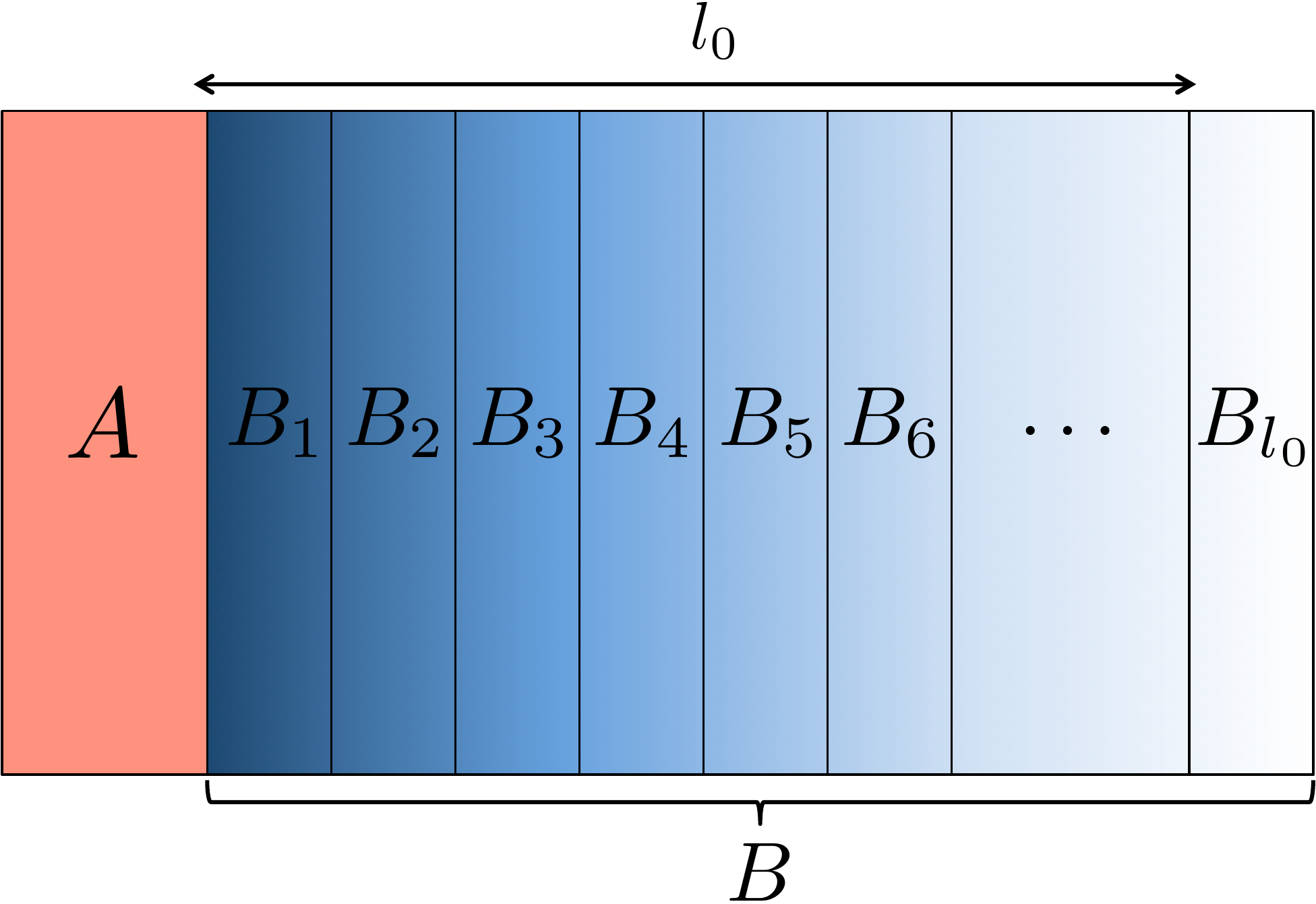}
}
\caption{(color online). Strengthening of the area law resulting from the Markov property. 
In the figure, we consider a 2D system and decompose it into $A$ and $B=B_1 B_2 \ldots B_{l_0}$ with $\dist_{A,B_l}=l$ ($1\le l\le l_0$).
}
\label{fig:strong_area_law}
\end{figure}


The second implication of the theorem is the strengthening of the area law.
The area law for mutual information was derived at arbitrary temperatures in Ref.~\cite{PhysRevLett.100.070502} in the following form:
\begin{align}\label{eq:arealaw}
\mi_\rho (A:B) \le c \beta |\partial A| , 
\end{align}
 where $AB=V$ and $c$ is an $\orderof{1}$ constant. 
The area law implies that $\mi_\rho(A:B')$ saturates as $B'\subset B$ grows to $B$; however,  
Eq.~\eqref{eq:arealaw} does not provide the saturation rate. 
Our result implies it saturates exponentially fast, and the mutual information between two subsystems is exponentially localized around the boundary between $A$ and $B$. 
For further understanding, let us decompose $B$ into $l_0$ slices, $B_1B_2 \ldots B_{l_0}$, with $\dist_{A,B_l}=l$ for $l=1,2,\ldots, l_0$ (see Fig.~\ref{fig:strong_area_law}). 
Then, the question is how rapidly the mutual information $\mi_\rho (A:B_1B_2 \cdots B_l)$ saturates to $\mi_\rho (A:B)$.
From the relation $\mi(A:C|B)=\mi(A:BC)-\mi(A:B)$ and Ineq.~\eqref{eq:quantum_approximate_markov_chain}, we have
\begin{align}
&\mi_\rho (A:B_1 \cdots B_l) - \mi_\rho (A:B_1 \cdots B_{l-1}) \notag \\
=&\mi_\rho (A:B_l| B_1 \cdots B_{l-1})  \sim (\beta/\beta_c)^{l/r},
\label{saturation_of_the area_law}
\end{align}
which shows exponential decay with respect to $l$.

\begin{figure}[tt]
\centering
{
\includegraphics[clip, scale=0.33]{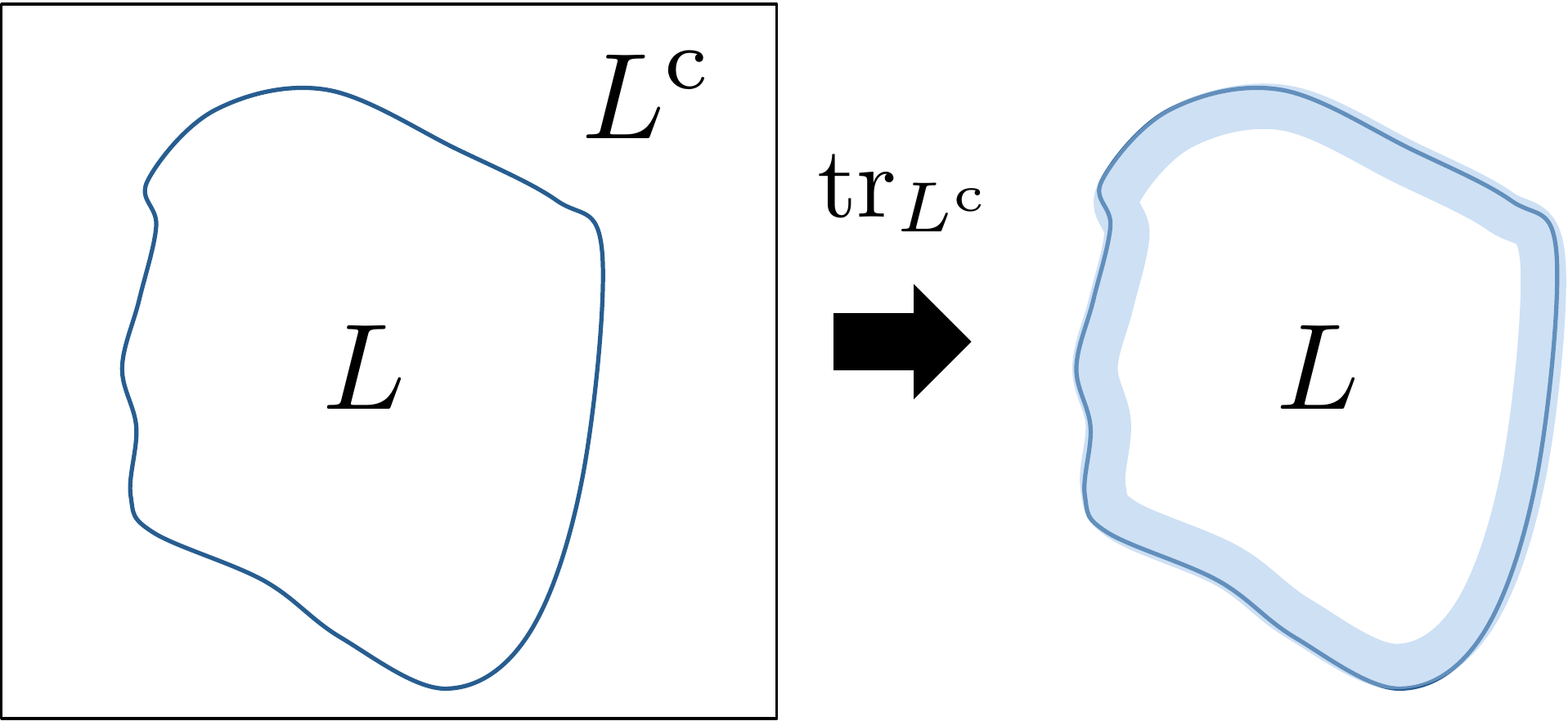}
}
\caption{(color online) Effective Hamiltonian $\tilde{H}_L$ for the reduced density matrix $\rho^L$. 
We decompose $\tilde{H}_L$  as $\tilde{H}_L = H_L + \Phi_L$, where $H_L$ is the original Hamiltonian in $L$ and $\Phi_L$ is the effective term that originates outside $L$. 
Theorem~\ref{main_lemma_effective_Ham} implies that $\Phi_L$ is exponentially localized around the surface region of $L$.
}
\label{result_outline}
\end{figure}

\sectionprl{Effective Hamiltonian on subsystem and classical simulation of Gibbs state} 
Theorem~\ref{main_theorem_QAMC_mutual_information} is related to the locality of the effective Hamiltonian.
We define the effective Hamiltonian of the local reduced density matrix as  
\begin{align}
\tilde{H}_L := -\beta^{-1} \log \tr_{L^\co} (e^{-\beta H}) . \label{def:tilde_H_L}
\end{align}
We formally describe $\tilde{H}_L$ as 
\begin{align}
\tilde{H}_L=H_L + \Phi_L , \label{effective_Hami_L_Phi}
\end{align}
where $H_L$ is composed of the original interacting terms in $H$ on subsystem $L$, namely, $H_L=\sum_{X\subset L} h_\ed$, and $\Phi_L$ is the effective interaction term.
We are interested in the locality of $\Phi_L$.  
Typically, it is computationally difficult to determine the effective term, even in classical Gibbs states~\cite{Foot3}.
Our present question is whether the (quasi-)locality of $\Phi_L$ can be ensured (Fig.~\ref{result_outline}).
In classical Gibbs states or systems with commuting Hamiltonians, $\Phi_L$ is exactly localized around the surface region of $L$ (not necessarily localized along the boundary). 
This point is crucial for the Gibbs states to be the exact Markov networks~\cite{brown2012quantum, Brandao2019}. 
Additionally, for systems with non-commuting Hamiltonians, the quasi-locality of $\Phi_L$ is numerically verified in Ref.~\cite{PhysRevB.81.054106}.
By following the same analysis as the proof of Theorem~\ref{main_theorem_QAMC_mutual_information}, we can rigorously prove the quasi-locality of $\Phi_L$ not only in the direction orthogonal to the boundary but also along the boundary.
\begin{theorem} \label{main_lemma_effective_Ham}
Using the setup and assumption of Theorem~\ref{main_theorem_QAMC_mutual_information}, $\Phi_L$ is approximated using a localized operator $\Phi_{\partial L_l}$ as follows:
\begin{align}
\| \Phi_L- \Phi_{\partial L_l}\| \le  \frac{e}{4\beta}  \frac{(\beta/\beta_c)^{l/r}}{1- \beta/\beta_c} |\partial L_r| ,\notag 
\end{align}
where $\Phi_{\partial L_l}$ is supported on the region $\partial L_l$ that has been defined in Eq.~\eqref{surface_region_of_L}. 
In addition, $\Phi_{\partial L_l}$ is composed of local operators that act on at most $(k\lfloor l/r \rfloor)$ spins (see supplementary materials~\cite{Supplement_Quantum_Markov} for an explicit form of $\Phi_{\partial L_l}$).
Moreover, computation of $\Phi_L$ up to a norm error of $n\epsilon$ is performed with the runtime bounded from above by 
\begin{align}
n(1/\epsilon)^{\orderof{k \log (d d_G^r )}},  \label{computational_cost_main}
\end{align}
where $d_G$ is the degree of the graph $G$.
\end{theorem}
This theorem immediately implies that the classical simulation of the Gibbs states is possible in polynomial time within an error of $1/\poly(n)$.
We note that the definition~\eqref{def:tilde_H_L} implies $\Phi_L=-\beta^{-1}\log (Z)$ for $L=\emptyset$; i.e., we can calculate the partition function using the same algorithm. We can also calculate the expectation values of local observables or the local entropy by explicitly obtaining the expression $\rho^L=e^{-\beta\tilde H_L}$. This is summarized in the following corollary.
\begin{corol} \label{corol:computation}
Thermodynamic properties such as local observables (e.g., energy and magnetization), the partition function $\log (Z)$, and local entropy $-\tr( \rho^L \log \rho^L)$ are classically simulated in polynomial time $\poly(n)$ as long as an error of $1/\poly(n)$ is allowed.
\end{corol}
\noindent
From Ref.~\cite{PhysRevB.91.045138}, we can prepare tensor network representations for arbitrary Gibbs states in the polynomial time of $n^{\orderof{\beta}}$. However, the classical simulation of the tensor network is \#P complete problem~\cite{PhysRevLett.98.140506,PhysRevResearch.2.013010} except in 1D cases. 
To the best of our knowledge, our result, for the first time, provides the fully polynomial-time approximation scheme (FPTAS~\cite{Foot7}) for the classical simulation of quantum Gibbs states, which is a quantum generalization of the FPATS for classical Gibbs states~\cite{
10.11451132516.1132538,
10.55552095116.2095191,
8104127}.


%
%
%
%
%
%
%
%
%

%

%
%
%

%
%
%
%

\sectionprl{Long-range interacting systems}
Finally, we extend Theorem~\ref{main_theorem_QAMC_mutual_information} from short-range interacting systems to long-range interacting systems.
We define the Hamiltonian with the power-law decay interaction assuming that $f(R)$ in \eqref{def:interaction_length} is given by
\begin{align}
f(R)= R^{-\alpha},  \label{long_range_interaction} 
\end{align}
where $\alpha>0$.
To consider a more general form as $f(R)= g R^{-\alpha}$, we must only scale the inverse temperature from $\beta$ to $\beta/g$. 
For example, we can consider the following Hamiltonian on a graph with a $D$-dimensional structure:
\begin{align}
H=\sum_{i,j \in V} \frac{J}{R_{i,j}^{\alpha+D}} h_{i,j} \with \|h_{i,j}\|=1, \notag
\end{align}
where $R_{i,j}$ is the distance between spins $i$ and $j$ defined by the graph structure $(V,E)$ and $J$ is determined so that 
inequality~\eqref{def:interaction_length} is satisfied.
This type of Hamiltonian is now controllable in realistic experiments and attracts considerable attention both in experimental~\cite{
yan2013observation,
richerme2014non,
jurcevic2014quasiparticle,
Islam583,
zhang2017observation} and theoretical aspects~\cite{
PhysRevLett.109.267203,
PhysRevLett.113.156402,
Kuwahara_2016_njp,
Kuwahara_2017,
kuwahara2019area}.


Similar to the case of short-range interacting systems, we prove the decay of the conditional mutual information for long-range interacting systems for $\alpha>0$.
\begin{theorem} \label{main_theorem_QAMC_mutual_information_long}
Let $A$, $B$, and $C$ be arbitrary subsystems in $V$ ($A,B,C \subset V$).
Then, under the assumptions of $\beta < \beta_c/11$ and $\dist_{A,C}\ge 2\alpha$,
the Gibbs state $\rho$ satisfies the approximate Markov property as follows:
\begin{align}
\mi_\rho(A:C|B)  \le \beta \min(|A|, |C|) \frac{C_{\beta} }{\dist_{A,C}^{\alpha}}   \label{eq:quantum_approximate_markov_chain_long},
\end{align}
where $C_{\beta}:=\frac{11e^{1/k}/\beta_c}{1-11\beta/\beta_c}$ and $\beta_c$ is as defined in \eqref{def:beta_c}.
\end{theorem}
By selecting $B=\emptyset$, we can also derive the power-law decay of the mutual information between two separated subsystems.
To the best of our knowledge, the clustering theorem for the Gibbs state with long-range interaction is limited for classical cases~\cite{Israel1979,Cammarota1982,Spohn1999,Procacci2001,KARGOL2005379,Menz2014} and special quantum cases~\cite{Kargol_2014,PhysRevLett.119.110601}. 
Our result provides the first general proof of the clustering theorem at finite temperatures in long-range interacting quantum systems. 

\sectionprl{Proof ideas of the main theorems}
We finally show the proof ideas to obtain the decay of the conditional mutual information.
The proof utilizes a high-temperature expansion. 
The difficulty lies in the fact that the standard cluster expansion technique cannot be applied to the logarithm of the reduced density matrix (e.g., $\rho^L$ with $L\subset \Lambda$). We introduce a new technique of the generalized cluster expansion, which allows us to systematically treat logarithmic operators (see Sec. I B in~\cite{Supplement_Quantum_Markov}).
Here, we parametrize the Hamiltonian~\eqref{eq:ham_graph} as  $H_{\vec{a}} = \sum_{X\in \Lambda} a_X h_X$ with $\vec{a}=\{a_X\}_{X\in \Lambda}$.
We then parametrize a target function of interest by $f_{\vec{a}}$ and directly expand it with respect to $\vec{a}$, where $f_{\vec{a}}$ can be chosen as  a scholar function and also as an operator function. Here, we choose the conditional mutual information as the function $f_{\vec{a}}$. 
The challenge in the generalized cluster expansion is to estimate the convergence radius of the expansion, where we need to consider a multiple derivative of the operators like $\log[\tr_{L^\co} (e^{-\beta H_{\vec{a}}})]$ with $L\subset \Lambda$. 
Our technical contributions are the systematical expression of the multiple derivative in the generalized cluster expansion (e.g., Propositions 3 and 4 in~\cite{Supplement_Quantum_Markov}) and the estimation of the convergence radius (see~\cite{Supplement_Quantum_Markov} for the details).

%



\sectionprl{Future perspective}
We here mention an open problem. 
The most important problem is the Markov property in low-temperature regimes, where our present analytical technique (i.e., the generalized cluster expansion~\cite{Supplement_Quantum_Markov}) breaks down.
It is no longer desirable that the Markov property holds for the arbitrary selections of the subregions $A$, $B$, and $C$ 
because the topological order can exist at finite temperatures in four-dimensional systems~\cite{PhysRevLett.107.210501}. 
Further, we hope to apply the current analyses to other essential problems, such as the contraction problem of the Projected Entangled Pair States~\cite{Lubasch_2014,PhysRevB.94.195143,PhysRevA.95.060102,Kastoryano2019} 
and efficiency guarantee of the heuristic classical algorithms for the quantum Gibbs states~\cite{PhysRevLett.93.207204,RevModPhys.73.33,PhysRevLett.97.187202,TANG2013557,PhysRevLett.102.190601}.   
  

\begin{acknowledgments}
We thank Keiji Saito for valuable discussions on this work.
The work of T. K. was supported by the RIKEN Center for AIP and JSPS KAKENHI Grant No. 18K13475. 
TK gives thanks to God for his wisdom.
K. K. acknowledges funding provided by the Institute for Quantum Information and
Matter, an NSF Physics Frontiers Center (NSF Grant {PHY}-{1733907}).
F. B. is supported by the NSF.

\sectionprl{Note added}
Regarding the classical simulations of quantum Gibbs states, we identified a related result obtained using a similar approach~\cite{Mehdi} at the same time of our submission.

\end{acknowledgments}

\bibliography{Markov}

\begin{thebibliography}{100}
\bibitem [{Foo({\natexlab{a}})}]{Foot2}%
  \BibitemOpen
  \href@noop {} {}For the cluster state on a ring with $n$ as an even integer,
  let $\rho^{V_0}$ be the reduced density matrix with respect to the odd spins,
  $V_0=\{1,3,5,\cdots,2s-1\}$ ($s:=n/2$). Then, we obtain the von Neumann
  entropy for $\rho^{V_0}$ as $S(\rho^{V_0})=(|V_0|-1) \log 2 = (s-1) \log 2$.
  On the contrary, for arbitrary subset $V'_0 \subset V_0$, the reduced density
  matrix, $\rho^{V_0'}$, is given by the identity matrix, and hence,
  $S(\rho_{V'_0}) = |V'_0| \log 2$. Therefore, for an arbitrary decomposition
  $V_0=ABC$, we have $\mi(A:C|B)=\log 2$ and the Markov property breaks
  down\BibitemShut {NoStop}%
\bibitem [{Foo({\natexlab{b}})}]{Foot3}%
  \BibitemOpen
  \href@noop {} {}When $\Phi_L$ is given for all $L\subset V$ with $|L|\le k$,
  we can compute the average energy $\tr(\rho H)$ in polynomial time. However,
  the computational complexity of calculating the average energy is known to be
  NP-hard~\cite{Barahona_1982}\BibitemShut {NoStop}%
\bibitem [{Foo({\natexlab{c}})}]{Foot7}%
  \BibitemOpen
  \href@noop {} {}Fully-Polynomial-Time Approximation Scheme (or FPTAS) is a
  type of deterministic algorithm which gives an approximation to a given
  problem up to an error $\epsilon$ with the runtime bounded from above by
  ${\rm poly}(n) \times {\rm poly}(1/\epsilon)$ with $n$ the input
  size.\BibitemShut {Stop}%
\end{thebibliography}

\appendix

\begin{widetext}

\section{Proof of Theorem~\ref{main_theorem_QAMC_mutual_information}}

\subsection{Preliminaries} 

We here recall the setup.
We consider a quantum spin system with $n$ spins, where each of the spin sits on a vertex of the graph $G=(V,E)$ with $V$ the total spin set ($|V|=n$).
For a partial set $L$ of spins, we denote the cardinality, that is, the number of vertices contained in $L$, by $|L|$ (e.g. $L=\{i_1,i_2,\ldots, i_{|L|}\}$).
We also denote the complementary subset of $L$ by $L^\co := V\setminus L$.
We denote the local Hilbert space by $\mathcal{H}^v$ ($v\in V$) with $\dim (\mathcal{H}^v)=d$ and the entire Hilbert space is given by $\mathcal{H}:=\bigotimes_{v\in V} \mathcal{H}^v$ with $\dim (\mathcal{H}) =d^n$.
We also define the local Hilbert space of the subset $L\subset V$ as $\mathcal{H}^{L}$ and denote the dimension by $d_L$, namely $d_L:=d^{|L|}$.
We define $\ban(\mathcal{H})$ as the space of bounded linear operators on $\mathcal{H}$.

When we consider a reduced operator on a subsystem $L$, we denote it as 
\begin{align}
O^L = \tr_{L^\co} (O)  \otimes \hat{1}_{L^\co} \in \ban(\mathcal{H})
\label{definition:reduced_matrix}
\end{align}
by using the superscript index, where $\hat{1}$ is the identity operator and $\tr_{L^\co}$ is the partial trace operation with respect to the Hilbert space $\mathcal{H}^{L^\co}$.

We also define the following set:
\begin{align}
E^{(x)} := \left\{X\subset V| \diam(X)=x, \quad  |X|\le k   \right\} \label{def:E_x_set}
\end{align}
with
\begin{align}
\diam(\ed):=\max_{v_1,v_2 \in \ed}\dist_{v_1,v_2},
\end{align}
where we defined $\dist_{A,B}$ as the shortest path length via $E$ which connects $A$ and $B$ ($A\subset V$, $B\subset V$).

In the setup of Theorem~\ref{main_theorem_QAMC_mutual_information}, we consider the Hamiltonian as
\begin{align}
H= \sum_{\ed\in E_r} h_\ed  ,\quad {\rm with} \quad
    \sum_{\ed|\ed\ni v} \|h_\ed\|  \le 1
    \for \forall v\in V
\label{eq:ham_graph_sup}
\end{align}
with 
\begin{align}
E_r:= E^{(1)} \sqcup E^{(2)} \sqcup \cdots \sqcup E^{(r)} \quad (r \in \mathbb{N}). \label{eq:ham_graph_sup2}
\end{align}
Here, the Hamiltonian~\eqref{eq:ham_graph_sup} describes an arbitrary $k$-body interacting systems with finite interaction length $r$.

Throughout the manuscript, we denote the natural logarithm by $\log (\cdot)$ for the simplicity, namely $\log (\cdot) = \log_e (\cdot)$.  

\subsubsection{Cluster notation}\label{subsub_sec:Preliminaries}

We then define several basic terminologies.
On the graph $(V,E)$, we call a multiset of subsystems $w=\{\ed_1,\ed_2,\ldots,\ed_{|w|}\}$ ($\ed_j\in E_r$ for $j=1,2,\ldots,|w|$) as ``cluster'', where $|w|$ is the cardinality of $w$.
Note that each of the elements $\{\ed_j\}_{j=1}^{|w|}$ satisfies $\diam(\ed_j)\le r$ from the definition~\eqref{eq:ham_graph_sup2}.
We denote $\Cl_{r,m}$ by the set of $w$ with $|w|=m$ and let $V_w \subseteq V$ and $E_w \subseteq E_r$ be the set of different vertices (or spins) and subsystems which are contained in $w$, respectively.
Also, we define connected clusters as follows:  

 \begin{definition}{\rm  \bf (Connected cluster)}\label{Def:Connected_cluster}
For a cluster $w\in \Cl_{r,|w|}$, we say that $w$ is a connected cluster if 
there are no decompositions of $w=w_1 \sqcup w_2$ such that $V_{w_1} \cap V_{w_2}= \emptyset$.
We denote by $\Gc_{r,m}$ the set of the connected clusters with $|w|=m$.
\end{definition}

 \begin{definition}{\rm  \bf (Connected cluster to a region, FIG.~\ref{fig:connected_cluster})}\label{Def:Connected_cluster_to_L}
Similarly, we say that $w\in \Cl_{r,|w|}$ is a connected cluster to a subsystem $L$ if
there are no decompositions of $w=w_1 \sqcup w_2$ such that $(L\cup V_{w_1}) \cap V_{w_2}= \emptyset$.
We denote by $\Gc_{r,m}^L$ the set of the connected clusters to $L$ with $|w|=m$. 
\end{definition}

 \begin{definition}{\rm  \bf (Connected cluster with a link between two regions, FIG.~\ref{fig:connected_cluster_A_B})}\label{Def:Connected_cluster_to_A_B}
Finally, for a connected cluster $w\in \Gc_{r,|w|}$, we say that $w$ has links between $A$ and $B$ if there exist a path from $A$ to $B$ in $E_w$. 
We denote by $\Gc_{r,m}^{A,B}$ the set of the connected clusters with $|w|=m$ which have a link $A$ and $B$. 
\end{definition}

\begin{figure}
\centering
\subfigure[Case of $w\in \Gc_{r,4}^L$]
{\includegraphics[clip, scale=0.42]{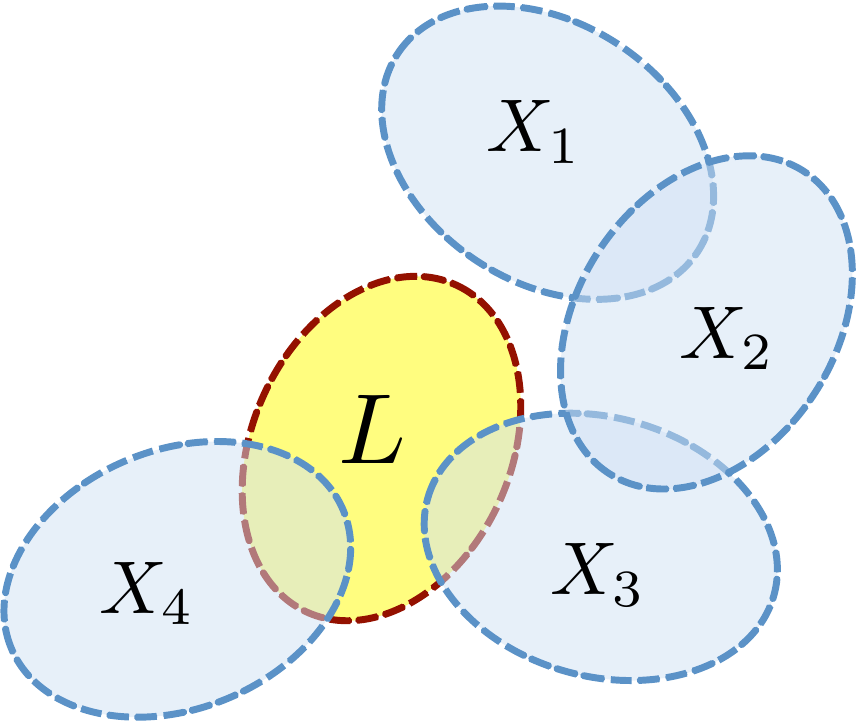}}
\subfigure[Case of $w\notin \Gc_{r,4}^L$]
{\includegraphics[clip, scale=0.42]{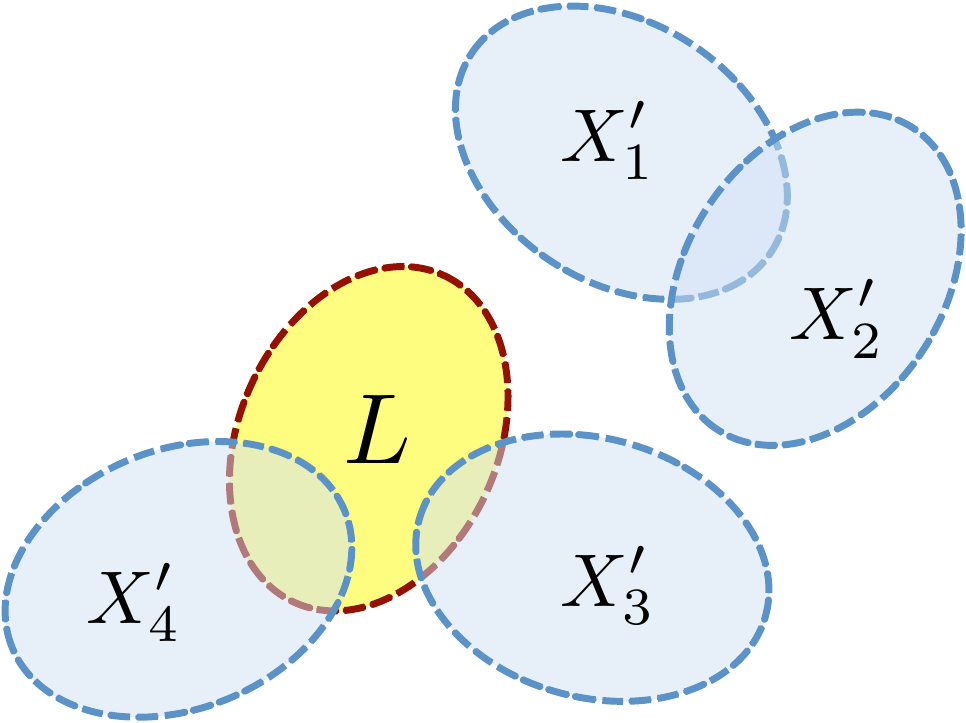}}
\caption{
Schematic pictures of clusters of $w \in \Gc_4^L$ and $w \notin \Gc_4^L$. Each of the elements $\{\ed_s | \ed_s \in E_r\}$ is a subset of the total set $V$ (i.e., $\ed \subset V$). In (a), there there are no decompositions of $w=w_1 \sqcup w_2$ such that $(L\cup V_{w_1}) \cap V_{w_2}= \emptyset$ for $w=\{\ed_1,\ed_2,\ed_3,\ed_4\}$, whereas  in (b) the decomposition $w'=w'_1 \sqcup w'_2$ with $w'_1=\{\ed'_2,\ed'_3\}$ and $w'_2=\{\ed'_1,\ed'_4\}$ satisfies $(L\cup V_{w_1}) \cap V_{w_2}= \emptyset$.
}
\label{fig:connected_cluster}
\end{figure}

\begin{figure}
\centering
\subfigure[Case of $w\in \Gc_{r,4}^{A,B}$
]
{\includegraphics[clip, scale=0.4]{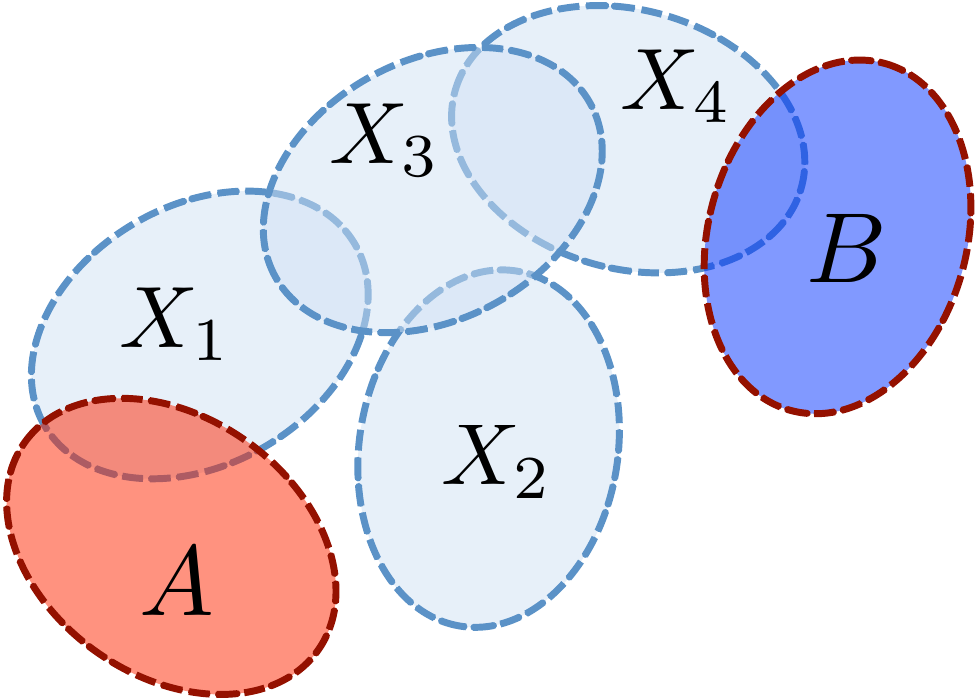}
}
\subfigure[Case of $w\notin \Gc_{r,4}^{A,B}$
]
{\includegraphics[clip, scale=0.4]{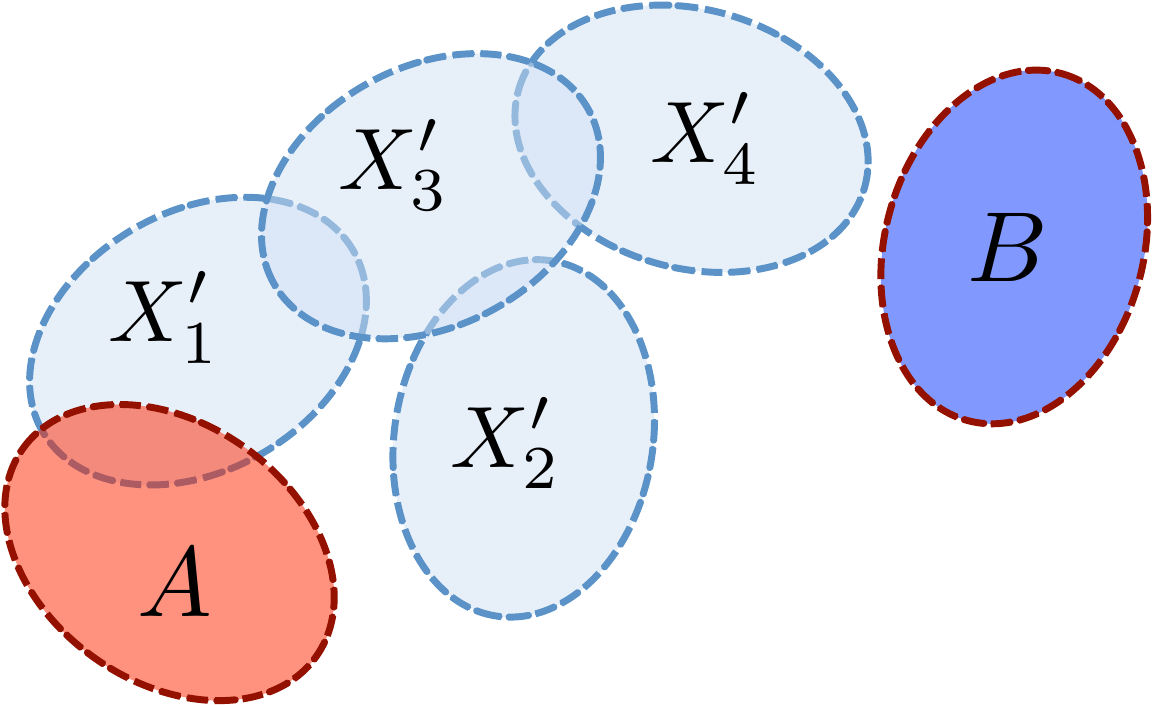}
}
\subfigure[Case of $w\notin \Gc_{r,4}^{A,B}$
]
{\includegraphics[clip, scale=0.4]{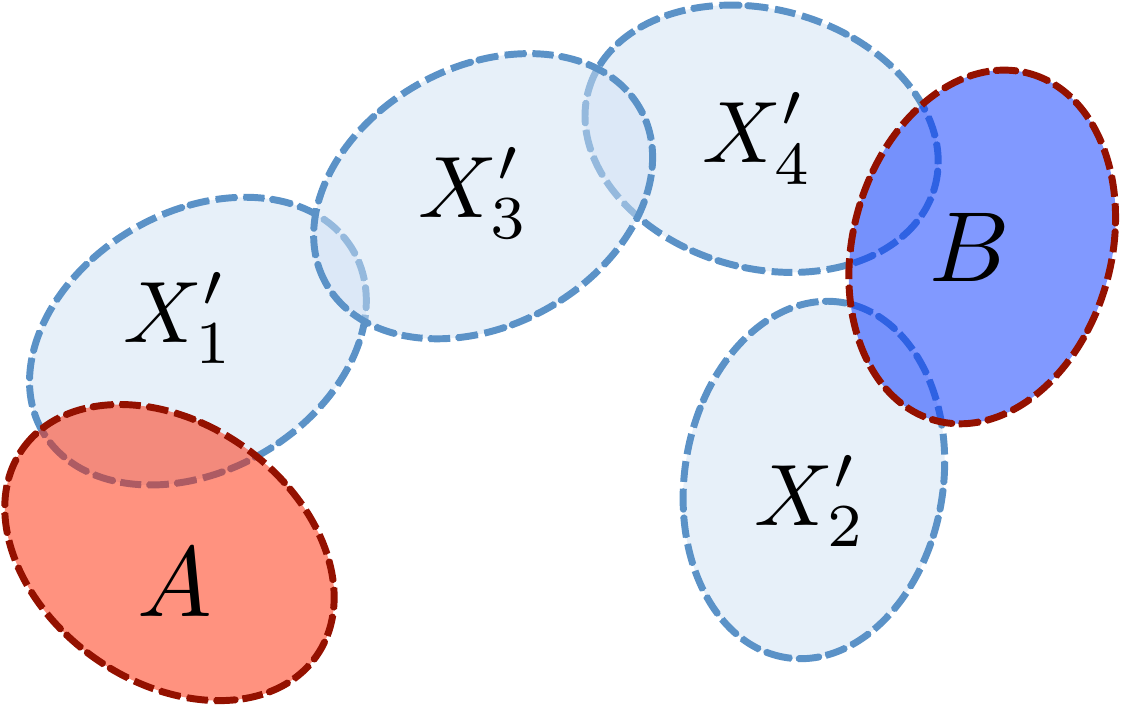}
}
\caption{Schematic pictures of clusters of $w \in \Gc_{r,4}^{A,B}$ and $w \notin \Gc_{r,4}^{A,B}$. In (a), subsystems $A$ and $B$ are connected with each other by the cluster $w$. On the other hand, in (b), the cluster $w$ does not have a link between $A$ and $B$, and in (c), the cluster has the link but is not connected.}
\label{fig:connected_cluster_A_B}
\end{figure}

\subsubsection{Basic lemmas for logarithmic operators}
 
Before going to the proof, we prove the following basic lemmas:
\begin{lemma} \label{lem:product_state}
Let $O\in \ban(\mathcal{H})$ be an arbitrary non-negative operator written as 
 \begin{align}
O = \Gamma_{L_1} \otimes \Gamma_{L_2}\otimes  \cdots \otimes \Gamma_{L_m},
\end{align}
where $\{ \Gamma_{L_j}\}_{j=1}^L \in \ban(\mathcal{H})$ are supported on the subsystems $\{L_j\}_{j=1}^m$, respectively and we assume $L_1 \sqcup L_2 \sqcup \cdots \sqcup L_m =V$. 
Then, for arbitrary subsystems $A,B,C\subset V$, we have
 \begin{align}
\log O^{AB} + \log O^{BC} -\log O^{ABC}- \log O^{B}   = \sum_{j=1}^m (\log  \Gamma_{L_j}^{AB} + \log  \Gamma_{L_j}^{BC} -\log  \Gamma_{L_j}^{ABC}- \log  \Gamma_{L_j}^{B}).
\label{decomp_product_state_cond}
\end{align}
Note that $\{O^{AB},O^{BC},O^{ABC},O^{B}\}$ are reduced operators as defined in Eq.~\eqref{definition:reduced_matrix}.
\end{lemma}

\textit{Proof of Lemma~\ref{lem:product_state}.}
We define $A_j := A \cap L_j$, $B_j := B \cap L_j$, and $C_j := C \cap L_j$ for $j=1,2,\ldots m$.
We notice that $\bigsqcup_{j=1}^m A_j = A$, $\bigsqcup_{j=1}^m B_j = B$ and $\bigsqcup_{j=1}^m C_j = C$ because of $\bigsqcup_{j=1}^m L_j =V$.
Then, from the definition~\eqref{definition:reduced_matrix}, the reduced operator of $O$ with respect to the subsystem $B$ is given by
 \begin{align}
O^{B} &= \tr_{L_1\setminus B_1}\left(\Gamma_{L_1}\right) \otimes \tr_{L_2\setminus B_2}\left(\Gamma_{L_2}\right) \otimes 
\cdots \otimes \tr_{L_m\setminus B_m}\left(\Gamma_{L_m}\right)  \otimes \hat{1}_{B^\co}\notag \\
&= \Gamma_{B_1} \otimes \Gamma_{B_2} \otimes \cdots \otimes  \Gamma_{B_m} \otimes \hat{1}_{B^\co} , \label{O_B_decomposition_lem}
\end{align}
where $\Gamma_{B_j} := \tr_{L_j\setminus B_j}\left(\Gamma_{L_j}\right) \otimes \hat{1}_{B_j^\co}$ for $j=1,2,\ldots , m$. 
We define $\Gamma_j^{BC}$, $\Gamma_j^{ABC}$ and  $\Gamma_j^{B}$ in the same way.
We thus obtain
 \begin{align}
\log O^{B} &= \sum_{j=1}^m \log  \Gamma_{B_j}. \label{O_B_decomposition_lem2}
\end{align}
On the other hand, we have from the definition~\eqref{definition:reduced_matrix} 
 \begin{align}
\Gamma_{L_j}^B= \tr_{B^\co} \left( \Gamma_{L_j} \right) \otimes \hat{1}_{B^\co} 
=d^{|B^\co| - |L_j \setminus B_j|}  \Gamma_{B_j} 
=d^{n-|B|-|L_j|+|B_j|} \Gamma_{B_j},
\end{align}
which reduces Eq.~\eqref{O_B_decomposition_lem2} to
 \begin{align}
\log O^{B} &=-(m-1)(n-|B|)\log (d) +  \sum_{j=1}^m \log \Gamma_{L_j}^B  . \label{O_B_decomposition_lem3}
\end{align}
We obtain the similar form to Eq.~\eqref{O_B_decomposition_lem3} for $O^{AB}$, $O^{BC}$ and $O^{ABC}$.
After a straightforward calculation, we prove the equation~\eqref{decomp_product_state_cond}. $\square$

Second, we prove the following lemma:
\begin{lemma} \label{lem:separate_state}
For an arbitrary non-negative operator $O\in \ban(\mathcal{H})$ which is given by the form of
 \begin{align}
O = O_{L} \otimes \hat{1}_{L^\co}
\end{align}
with $L\cap C=\emptyset$, we have 
 \begin{align}
\log O^{AB} + \log O^{BC} -\log O^{ABC}- \log O^{B}  = 0. \label{O_AB_BC_ABC_B_eq_0}
\end{align}
\end{lemma}

\textit{Proof of Lemma~\ref{lem:separate_state}.}
From the definition, we obtain  
 \begin{align}
O^{ABC} = O^{AB} \otimes \hat{1}_{C}. 
\end{align}
Thus, we obtain $\log O^{BC}=\log (  O^B \otimes \hat{1}_{C}) $ and $\log O^{ABC} =\log O^{AB}$, and hence we immediately obtain Eq.~\eqref{O_AB_BC_ABC_B_eq_0}. This completes the proof.
$\square$


\subsection{Generalized cluster Expansion} \label{cluster_expansion_cond_mutual_information}

We first parametrize $H$ by using a parameter set $\vec{a}:=\{a_\ed\}_{\ed\in E_r}$ as 
\begin{align}
H_{\vec{a}} = \sum_{\ed\in E_r} a_\ed h_\ed, \label{eq_parameterize_H}
\end{align}
where $ H=H_{\vec{1}}$ with $\vec{1}=\{1,1,\ldots,1\}$.
Note that there are $|E_r|$ parameters in total. 
By using Eq.~\eqref{eq_parameterize_H}, we define a parametrized Gibbs state $\rho_{\vec{a}}$ as  
\begin{align}
\rho_{\vec{a}} := \frac{e^{-\beta H_{\vec{a}}}}{Z_{\vec{a}}} ,
\end{align}
where $Z_{\vec{a}}:= \tr(e^{-\beta H_{\vec{a}}})$.

In the standard cluster expansion, we consider the Taylor expansion of $e^{-\beta H_{\vec{a}}}$ with respect to the parameters $\vec{a}$.
It works well in analyzing a correlation function or tensor network representation, while it is not appropriate to analyze the entropy or effective Hamiltonian of a reduced density matrix.
To overcome it, we generalize the standard cluster expansion.
We parametrize a target function of interest by $f_{\vec{a}}$ and directly expand it with respect to $\vec{a}$, where $f_{\vec{a}}$ can be chosen not only as  a scholar function but also as a operator function.
Here, we choose the conditional mutual information as the function $f_{\vec{a}}$. 
By using $\rho_{\vec{a}}$, we parameterize the conditional mutual information by $\mi_{\vec{a}}(A:C|B)$ in the following form:
\begin{align}
\mi_{\vec{a}}(A:C|B) &= - \tr \left[\rho \left(\log \rho^{AB}_{\vec{a}} + \log \rho^{BC}_{\vec{a}} -\log \rho^{ABC}_{\vec{a}} - \log \rho^{B}_{\vec{a}}\right) \right] \notag \\
&=- \tr \left[\rho \left(\log \tilde{\rho}^{AB}_{\vec{a}} + \log \tilde{\rho}^{BC}_{\vec{a}} -\log \tilde{\rho}^{ABC}_{\vec{a}} - \log \tilde{\rho}^{B}_{\vec{a}}\right) \right] ,
\label{cond_mutual_info_tilde_rho}
\end{align}
where $\rho=\rho_{\vec{1}}$ and we define $\tilde{\rho}_{\vec{a}}$ as 
\begin{align}
\tilde{\rho}_{\vec{a}} :=  e^{-\beta H_{\vec{a}}}   \label{def_rho_L_tilde_rho_a}
\end{align}
with
\begin{align}
\tilde{\rho}^{L}_{\vec{a}} =  \left( e^{-\beta H_{\vec{a}}} \right)^L =  \tr_{L^\co} \left( e^{-\beta H_{\vec{a}}} \right) \otimes \hat{1}_{L^\co} . \label{def_rho_L_tilde_rho_L}
\end{align}
Note that we use the definition~\eqref{definition:reduced_matrix} for  $\tilde{\rho}_{\vec{a}}^{L}$ ($L\subset V$)

In the following, we define 
\begin{align}
\tilde{H}_{\vec{a}}(A:C|B) :=  \log \tilde{\rho}^{AB}_{\vec{a}} + \log \tilde{\rho}^{BC}_{\vec{a}} -\log \tilde{\rho}^{ABC}_{\vec{a}} - \log \tilde{\rho}^{B}_{\vec{a}} ,
\label{def_tilde_H_a_A_C_B}
\end{align}
which gives 
\begin{align}
\mi_{\vec{a}}(A:C|B) = \tr \left[\rho \tilde{H}_{\vec{a}}(A:C|B)\right] \le \|\tilde{H}_{\vec{a}}(A:C|B)\| .\label{cond_mutual_info_tilde_rho_upper_bound}
\end{align}

Then, the Taylor expansion with respect to $\vec{a}$ to the operator $\tilde{H}_{\vec{a}}(A:C|B)$ reads 
\begin{align}
\tilde{H}_{\vec{1}}(A:C|B)  =\sum_{m=0}^{\infty}\frac{1}{m!} \left[  \left( \sum_{\ed \in E_r}  \frac{\partial}{\partial a_\ed}  \right)^m \tilde{H}_{\vec{a}} (A:C|B) \right]_{\vec{a}=\vec{0}} ,
\end{align}
where $\vec{0}=\{0,0,\ldots,0\}$. 
By using the cluster notation, we obtain 
\begin{align}
\sum_{\ed_1,\ed_2,\ldots,\ed_m\in  E_r} = \sum_{w \in \Cl_{r,m}} n_w ,
\end{align}
which yields
\begin{align}
\tilde{H}_{\vec{1}}(A:C|B)  =& \sum_{m=1}^{\infty}\frac{1}{m!} \sum_{\ed_1,\ed_2,\ldots,\ed_m\in E_r} \prod_{j=1}^m \frac{\partial}{\partial a_{\ed_j}}\tilde{H}_{\vec{a}}(A:C|B)  \Bigl|_{\vec{a}=\vec{0}}= \sum_{m=1}^{\infty}\frac{1}{m!} \sum_{w \in \Cl_{r,m}} n_w \Der_{w}  \tilde{H}_{\vec{a}}(A:C|B)\Bigl|_{\vec{a}=\vec{0}}, \label{Cluster_decomp_mutual}
\end{align}
where $w=\{\ed_1, \ed_2 \ldots, \ed_m\}$ and $n_w$ is the multiplicity that $w$ appears in the summation, and we defined
\begin{align}
\Der_{w}:= \prod_{j=1}^m \frac{\partial}{\partial a_{\ed_j}} \with w=\{\ed_1, \ed_2 \ldots, \ed_m\}.
\end{align}
We notice that the partial derivatives $\frac{\partial}{\partial a_{\ed}}$ and $\frac{\partial}{\partial a_{\ed'}}$ commute with each other because $\log (\tilde{\rho}^{L}_{\vec{a}})$ is a $C^\infty$-smooth function with respect to $\vec{a}$ as long as the system size $n$ is finite.
The $C^\infty$-smoothness of $\log (\tilde{\rho}^{L}_{\vec{a}})$ is proved as follows:
For a finite system size $n$, the $C^\infty$-smoothness of $e^{-\beta H_{\vec{a}}}$ is ensured, and hence $\tilde{\rho}^{L}_{\vec{a}}$ is also $C^\infty$-smooth from the definition~\eqref{def_rho_L_tilde_rho_L}.
Also, we can set 
\begin{align}
\| e^{- \tau \hat{1}} \tilde{\rho}^{L}_{\vec{a}}\| \le 1.
\end{align}
by choosing a finite energy $\tau<\infty$ appropriately.
Notice that $e^{- \tau\hat{1}} \tilde{\rho}^{L}_{\vec{a}}$ is Hermitian and $e^{- \tau\hat{1}}\tilde{\rho}^{L}_{\vec{a}} \succeq 0$. 
This implies the absolute convergence of the following expansion:
\begin{align}
\log ( \tilde{\rho}^{L}_{\vec{a}}) &= \tau\hat{1} + \log (e^{-\tau \hat{1}} \tilde{\rho}^{L}_{\vec{a}}) =  \tau\hat{1}+  \log (\hat{1} +  e^{- \tau \hat{1}} \tilde{\rho}^{L}_{\vec{a}} -\hat{1}) = \tau\hat{1}+ \sum_{m=1}^\infty \frac{(-1)^{m-1}}{m} (e^{- \tau \hat{1}}\tilde{\rho}^{L}_{\vec{a}} -\hat{1})^m.
\end{align}
Thus, the $C^\infty$-smoothness of $\tilde{\rho}^{L}_{\vec{a}}$ implies of $C^\infty$-smoothness of $\log (\tilde{\rho}^{L}_{\vec{a}})$.

Note that the case of $m=0$ (i.e., $|w|=0$) does not contribute to the expansion because of $\tilde{H}_{\vec{0}} (A:C|B)=0$.
In order to calculate the summation of $\sum_{w \in \Cl_{r,m}}$, we utilize the following proposition:
\begin{prop} \label{prop: connceted_cluster_mutual}
The cluster expansion~\eqref{Cluster_decomp_mutual} reduces to the summation of connected clusters which have links between $A$ and $C$:
\begin{align}
\tilde{H}_{\vec{1}}(A:C|B)  =\sum_{m=1}^{\infty}\frac{1}{m!} \sum_{w \in \Gc^{A,C}_{r,m}} n_w \Der_{w}  \tilde{H}_{\vec{a}}(A:C|B)\Bigl|_{\vec{a}=\vec{0}}, \label{generalized_cluster_exp_tilde_H}
\end{align}
where the definition of $\Gc^{A,C}_{r,m}$ has been given in Def.~\ref{Def:Connected_cluster_to_A_B}. 
\end{prop}
From this proposition, we only need to estimate the contribution of clusters in $\Gc^{A,C}_{r,m}$ to upper-bound the conditional mutual information $\mi_{\vec{1}}(A:C|B)=\tr[\rho \tilde{H}_{\vec{1}}(A:C|B)]$.

\subsubsection{Proof of Proposition~\ref{prop: connceted_cluster_mutual}} \label{proof_connected_cluster_contribute}

We first introduce the notation $\vec{a}_w$ as a parameter vector such that the elements $\{a_\ed\}_{\ed\notin w}$ are vanishing, that is,
\begin{align}
( \vec{a}_{w} )_{\ed} =0 \for  \ed \notin  w , \label{def:a_w_para}
\end{align}
where we denote an element of $a_{\ed}$ in $\vec{a}$ by $( \vec{a})_{\ed}$.
We then obtain 
\begin{align}
\Der_{w}  \tilde{H}_{\vec{a}}(A:C|B)  \Bigl|_{\vec{a}=\vec{0}} = \Der_{w} \tilde{H}_{\vec{a}_w}(A:C|B) \Bigl|_{\vec{a}_{w}=\vec{0}} .
\end{align}

In the following, we aim to prove 
\begin{align}
 \Der_{w}\tilde{H}_{\vec{a}_{w}}(A:C|B)  \Bigl|_{\vec{a}_{w}=\vec{0}}  =0 \for  w \notin \Gc^{A,C}_{r,|w|}.
\end{align}
We notice that if $w \notin \Gc^{A,C}_{r,|w|}$ the cluster $w$ satisfies either one of the following two properties (see Figs.~\ref{fig:connected_cluster_A_B} (b) and (c)): 
\begin{align}
L_{w} \cap A = \emptyset \quad {\rm or} \quad L_{w} \cap C= \emptyset \label{first_case_QAMC_AC}
\end{align}
and
\begin{align}
w \notin \Gc_{r,|w|}\label{second_case_QAMC_AC}.
\end{align}

In the first case~\eqref{first_case_QAMC_AC},  we can immediately obtain $\tilde{H}_{\vec{a}_{w}}(A:C|B)=0$ by choosing $O=e^{-\beta H_{\vec{a}_{w}}}$ in the lemma~\ref{lem:separate_state}. 
In the second case~\eqref{second_case_QAMC_AC}, there exists a decomposition of $w=w_1\sqcup w_2$ ($|w_1|, |w_2|>0$) such that $V_{w_1} \cap V_{w_2} = \emptyset$.
Hence, we have $e^{-\beta H_{\vec{a}_{w}}} =e^{-\beta H_{\vec{a}_{w_1}}} \otimes e^{-\beta H_{\vec{a}_{w_2}}}$, and from Lemma~\ref{lem:product_state} we obtain
\begin{align}
\tilde{H}_{\vec{a}_w}(A:C|B) = \tilde{H}_{\vec{a}_{w_1}}(A:C|B) +  \tilde{H}_{\vec{a}_{w_2}}(A:C|B) .
\end{align}
Because of $\Der_{w_2} \tilde{H}_{\vec{a}_{w_1}}(A:C|B)= \Der_{w_1} \tilde{H}_{\vec{a}_{w_2}}(A:C|B)=0$, we have $\Der_w \tilde{H}_{\vec{a}_w}(A:C|B)=0$.
This completes the proof of Proposition~\ref{prop: connceted_cluster_mutual}. $\square$

 {~}

\hrulefill{\bf [ End of Proof of Proposition~\ref{prop: connceted_cluster_mutual} ] }

{~}


\subsection{Estimation of the expanded terms}


In order to estimate the summation~\eqref{generalized_cluster_exp_tilde_H} with respect to $\sum_{w \in \Gc^{A,C}_{r,m}}$,  we consider a derivative of 
 \begin{align}
\Der_w \log \tilde{\rho}^{L}_{\vec{a}}  \Bigl|_{\vec{a}=\vec{0}}= \Der_w  \log \tilde{\rho}^{L}_{\vec{a}_w}  \Bigl|_{\vec{a}_w=\vec{0}}
\label{Der_w_vec_a_vec_a_w}
\end{align}
for an arbitrary subsystem $L\subset V$.
We choose the subsets $AB$, $BC$, $ABC$ and $B$ as $L$ afterward.
We here give an explicit form of the derivative $\Der_w\log \tilde{\rho}^{L}_{\vec{a}}$ in the following proposition~\ref{prop:explicit_form_log_derivation}.

\begin{prop} \label{prop:explicit_form_log_derivation}
Let us take $m-1$ copies of the partial Hilbert space $\mathcal{H}^{L^\co}$ and distinguish them by $\{\mathcal{H}_j^{L^\co}\}_{j=1}^m$. 
Then, we define the extended Hilbert space  as $\mathcal{H}^L \otimes \mathcal{H}^{L^\co} _{1:m}$ with
\begin{align}
\mathcal{H}^{L^\co} _{1:m}:=\mathcal{H}^{L^\co}_{1} \otimes \mathcal{H}^{L^\co}_{2} \otimes \cdots \otimes\mathcal{H}^{L^\co}_{m}.  \label{copy_hilbert_space_log}
\end{align}
Then, for an arbitrary operator $O\in\mathcal{H}$, 
we extend the domain of definition and denote $O_{\tilde{\mathcal{H}}_s}\in \ban( \mathcal{H}^L \otimes \mathcal{H}^{L^\co} _{1:m})$ by the operator which acts only on the space $\mathcal{H}^L \otimes \mathcal{H}_s^{L^\co}$.
Now, for an arbitrary cluster $w=\{\ed_1,\ed_2,\ldots,\ed_m\}$, we have 
\begin{align}
 &\Der_{w}\log \tilde{\rho}^{L}_{\vec{a}} \bigl|_{\vec{a}=\vec{0}} =\frac{(-\beta)^m}{m! d_{L^\co}^{m}} \mathcal{P}_m  \tr_{L^\co_{1:m}} \left (\tilde{h}_{\ed_1}^{(0)} \tilde{h}_{\ed_2}^{(1)}\cdots \tilde{h}_{\ed_m}^{(m-1)} \right),  \label{simple_expression_der_w_log}
\end{align}
where $\tr_{L^\co_{1:m}}$ denotes the partial trace with respect to the Hilbert space $\mathcal{H}^{L^\co} _{1:m}$ and we define
\begin{align}
\tilde{O}^{(0)} :=O_{\tilde{\mathcal{H}}_1} ,\quad \tilde{O}^{(s)} := O_{\tilde{\mathcal{H}}_1} + O_{\tilde{\mathcal{H}}_2} + \cdots +O_{\tilde{\mathcal{H}}_{s}} -  s O_{\tilde{\mathcal{H}}_{s+1}} \label{Def:tilde_O_s}
\end{align}
for $s=1,2, \ldots,m$.
Note that $\mathcal{P}_m$ is the symmetrization operator as
\begin{align}
 \mathcal{P}_m  \tilde{h}_{\ed_1}^{(0)} \tilde{h}_{\ed_2}^{(1)}\cdots \tilde{h}_{\ed_m}^{(m-1)}
 =\sum_{\sigma} \tilde{h}_{\ed_{\sigma_1}}^{(0)} \tilde{h}_{\ed_{\sigma_2}}^{(1)} \cdots \tilde{h}_{\ed_{\sigma_m}}^{(m-1)} ,  \label{Def:math_P_m}
\end{align}
where $\sum_\sigma$ denotes the summation of $m!$ terms which come from all the permutations. 
\end{prop}

\subsubsection{Proof of Proposition~\ref{prop:explicit_form_log_derivation}}
For the proof, we consider the Taylor expansion with respect to $\beta$:
\begin{align}
\log \tilde{\rho}^{L}_{\vec{a}}  = \sum_{m=0}^{\infty} \frac{\beta^{m}}{m!}
\frac{\partial^m}{\partial \beta^m} \log \tilde{\rho}^{L}_{\vec{a}} \biggl |_{\beta=0} .\label{taylor_exp_omega_ed_phi}
\end{align}
Next, because of 
\begin{align}
\frac{\partial^m}{\partial \beta^m} \log (d_{L^\co}) =0 \for m\ge 1,
\end{align}
we have 
\begin{align}
\frac{\partial^m}{\partial \beta^m} \log \tilde{\rho}^{L}_{\vec{a}} \Bigl|_{\beta=0}= \frac{\partial^m}{\partial \beta^m} \log \left[ \tr_{L^\co}  (e^{-\beta H_{\vec{a}}}/d_{L^\co}) \right] \Bigl|_{\beta=0} 
\end{align}
for $m\ge 1$. 

We aim to prove the following lemma which gives the explicit form of the derivatives with respect to $\beta$:
\begin{lemma} \label{thm:express_Der_operator}
The derivatives of $\log \tilde{\rho}^{L}_{\vec{a}}$ 
with respect to $\beta$  can be written  as
\begin{align}
 \frac{\partial^m}{\partial \beta^m} \log \left[ \tr_{L^\co}  (e^{-\beta H_{\vec{a}}}/d_{L^\co}) \right] \Bigl|_{\beta=0} &=\frac{(-1)^{m}}{d_{L^\co}^{m}}  \tr_{L_{1:m}^\co}\left( \tilde{H}_{\vec{a}}^{(0)}\tilde{H}_{\vec{a}}^{(1)}\cdots \tilde{H}_{\vec{a}}^{(m-1)}  \right), \label{derivative_log_beta_general}
\end{align}
where the definitions of $\tilde{H}_{\vec{a}}^{(s)}$ ($s=0,1,2,\ldots, m-1$) and $\mathcal{H}^{L^\co} _{1:m}$ have been given in Eqs.~\eqref{Def:tilde_O_s} and Eq.~\eqref{copy_hilbert_space_log}, respectively.
We give the proof of the lemma afterward.
\end{lemma}

By assuming the above lemma, we can prove Eq.~\eqref{simple_expression_der_w_log} as follows.
In considering $\Der_{w} \log \tilde{\rho}^{L}_{\vec{a}} \bigl|_{\vec{a}=\vec{0}} $ with $|w|=m$, only the $m$th order terms of $\beta$
in the expansion~\eqref{taylor_exp_omega_ed_phi} contribute to the derivative. 
Hence, we have
\begin{align}
\Der_{w} \log \tilde{\rho}^{L}_{\vec{a}}\bigl|_{\vec{a}=0}
 = \frac{\beta^{m}}{m!}\Der_{w} \left(  \frac{\partial^m}{\partial \beta^m} \log \left[ \tr_{L^\co}  (e^{-\beta H_{\vec{a}}}/d_{L^\co}) \right]  \right)  \biggl|_{\beta=0}.
 \label{derivative_log_beta_w_general}
\end{align}
By combining Eqs.~\eqref{Der_w_vec_a_vec_a_w}, \eqref{derivative_log_beta_general} and \eqref{derivative_log_beta_w_general}, we have
 \begin{align}
\Der_{w} \log \tilde{\rho}^{L}_{\vec{a}}\bigl|_{\vec{a}=0}&=  \frac{(-\beta)^{m}}{m!}\frac{1}{d_{L^\co}^{m}}  \Der_{w}\tr_{L_{1:m}^\co}\left( \tilde{H}_{\vec{a}}^{(0)}\tilde{H}_{\vec{a}}^{(1)}\cdots \tilde{H}_{\vec{a}}^{(m-1)}  \right)   \notag \\
&= \frac{(-\beta)^{m}}{m!} \frac{1}{d_{L^\co}^{m}}  \mathcal{P}_m  \tr_{L^\co_{1:m}} \left (\tilde{h}_{\ed_1}^{(0)} \tilde{h}_{\ed_2}^{(1)}\cdots \tilde{h}_{\ed_m}^{(m-1)} \right).
\end{align}
We therefore obtain Eq.~\eqref{simple_expression_der_w_log} in Proposition~\ref{prop:explicit_form_log_derivation}.
This completes the proof. $\square$

{~}

{~}

{\bf [Proof of Lemma~\ref{thm:express_Der_operator}]} 
In order to prove Eq.~\eqref{derivative_log_beta_general}, we first expand $ \log \left[ \tr_{L^\co} (e^{-\beta H_{\vec{a}}}/d_{L^\co}) \right]$ as follows:
\begin{align}
 \log \left[\frac{\tr_{L^\co} (e^{-\beta H_{\vec{a}}})}{d_{L^\co}}  \right]&=  \log \left[ \hat{1} + \sum_{m=1}^\infty \frac{(-\beta)^m}{m!} 
 \frac{\tr_{L^\co} (H_{\vec{a}}^m)}{d_{L^\co}}    \right] =\sum_{q=1}^\infty \frac{(-1)^{q-1}}{q}  \left(\sum_{m=1}^\infty \frac{(-\beta)^m}{m!} \frac{\tr_{L^\co} (H_{\vec{a}}^m)}{d_{L^\co}}\right)^q,
 \label{taylor_expansion_log_e_lco}
\end{align}
where in the first equation we use the fact that $0$th term of the expansion gives $\tr_{L^\co} (\hat{1}/d_{L^\co}) =\hat{1}$.    
We then pick up the terms of $\beta^m$. 
Because of 
\begin{align}
&\left(\sum_{m=1}^\infty \frac{(-\beta)^m}{m!} \frac{\tr_{L^\co} (H_{\vec{a}}^m)}{d_{L^\co}}\right)^q\notag\\
=&\sum_{m=q}^\infty \sum_{\substack{m_1+m_2+\cdots+ m_q=m\\ m_1\ge1,m_2\ge1,\ldots,m_q\ge1}} \frac{(-\beta)^{m_1+m_2+\cdots+m_q}}{m_1!m_2!\cdots m_q!} 
\frac{ \tr_{L^\co}(H_{\vec{a}}^{m_1}) \tr_{L^\co}(H_{\vec{a}}^{m_2})   \cdots \tr_{L^\co}(H_{\vec{a}}^{m_q})}{d_{L^\co}^q}  ,
\end{align}
the $m$th-order term in Eq.~\eqref{taylor_expansion_log_e_lco} is given by
\begin{align}
\beta^m \sum_{q=1}^m\frac{(-1)^{q-1}}{q}\sum_{\substack{m_1+m_2+\cdots+ m_q=m\\ m_1\ge1,m_2\ge1,\ldots,m_q\ge1} } \frac{(-1)^{m}}{m_1!m_2!\cdots m_q!} 
\frac{\tr_{L^\co}(H_{\vec{a}}^{m_1}) \tr_{L^\co}(H_{\vec{a}}^{m_2})   \cdots \tr_{L^\co}(H_{\vec{a}}^{m_q})}{d_{L^\co}^q} .  \label{beta_m_term_complex0}
\end{align}
We thus obtain
\begin{align}
& \frac{\partial^m}{\partial \beta^m} \log \left[ \tr_{L^\co}  (e^{-\beta H_{\vec{a}}}/d_{L^\co}) \right] \Bigl|_{\beta=0}  \notag \\
 =& \sum_{q=1}^m\frac{(-1)^{q-1}}{q}\sum_{\substack{m_1+m_2+\cdots+ m_q=m\\ m_1\ge1,m_2\ge1,\ldots,m_q\ge1} } \frac{m! (-1)^{m}}{m_1!m_2!\cdots m_q!} 
\frac{\mathcal{P}_q \tr_{L^\co}(H_{\vec{a}}^{m_1}) \tr_{L^\co}(H_{\vec{a}}^{m_2})   \cdots \tr_{L^\co}(H_{\vec{a}}^{m_q})}{q!d_{L^\co}^q}  ,\label{beta_m_term_complex}
\end{align}
where $\mathcal{P}_q$ is the symmetrization operator with respect to $\{m_1,m_2,\ldots,m_q\}$.
In the same manner, we can formally expand 
\begin{align}
&\frac{(-1)^{m}}{d_{L^\co}^{m}}  \tr_{L_{1:m}^\co}\left( \tilde{H}_{\vec{a}}^{(0)}\tilde{H}_{\vec{a}}^{(1)}\cdots \tilde{H}_{\vec{a}}^{(m-1)}  \right)  \notag \\
=&\sum_{q=1}^m \sum_{\substack{m_1+m_2+\cdots+ m_q=m\\ m_1\ge1,m_2\ge1,\ldots,m_q\ge1} } \mathcal{C}^{(q)}_{m_1,m_2,\ldots,m_q} \mathcal{P}_q \tr_{L^\co}(H_{\vec{a}}^{m_1}) \tr_{L^\co}(H_{\vec{a}}^{m_2})   \cdots \tr_{L^\co}(H_{\vec{a}}^{m_q}) .
\label{derivative_log_beta_general_proof}
\end{align}

For the proof of Lemma~\ref{thm:express_Der_operator}, 
we need to check whether each of the coefficients of $\mathcal{P}_q \tr_{L^\co}(H_{\vec{a}}^{m_1}) \tr_{L^\co}(H_{\vec{a}}^{m_2})   \cdots \tr_{L^\co}(H_{\vec{a}}^{m_q}) $ for all the pairs of $\{m_1,m_2,\ldots,m_q\}$ is equal between Eqs.~\eqref{beta_m_term_complex} and \eqref{derivative_log_beta_general_proof}.
Instead of directly writing down the explicit form of $\mathcal{C}^{(q)}_{m_1,m_2,\ldots,m_q}$, we will take the following step.
First, we prove 
\begin{align}
\frac{\partial^m}{\partial \beta^m} \log \left[ \tr_{L^\co}  (e^{-\beta H_{\vec{a}}}/d_{L^\co}) \right] \Bigl|_{\beta=0}  
=\frac{(-1)^{m}}{d_{L^\co}^{m}}  \tr_{L_{1:m}^\co}\left( \tilde{H}_{\vec{a}}^{(0)}\tilde{H}_{\vec{a}}^{(1)}\cdots \tilde{H}_{\vec{a}}^{(m-1)}  \right) 
\label{in_the_case_of_L_co_eq_V}
\end{align}
in the case of $L^\co=V$. 
The proof of Eq.~\eqref{in_the_case_of_L_co_eq_V} implies that the coefficients of $\mathcal{P}_q \tr_{L^\co}(H_{\vec{a}}^{m_1}) \tr_{L^\co}(H_{\vec{a}}^{m_2})   \cdots \tr_{L^\co}(H_{\vec{a}}^{m_q}) $ are equal between Eqs.~\eqref{beta_m_term_complex} and \eqref{derivative_log_beta_general_proof} for $L^\co=V$. 
Then, because the coefficients $\mathcal{C}^{(q)}_{m_1,m_2,\ldots,m_q}$ do not depend on the form of $L^\co$, the proof in the case of $L^\co=V$ also results in the 
proof in the other cases (i.e., $L^\co \neq V$). 
Therefore, in the following, we aim to give the proof of Eq.~\eqref{in_the_case_of_L_co_eq_V} for $L^\co=V$.

For $L^\co=V$, we have
\begin{align}
 \frac{\partial}{\partial \beta}  \log \left[\frac{\tr_{V} (e^{-\beta H_{\vec{a}}})}{d_{V}}  \right]  &=- \tr (  H_{\vec{a}} \rho_{\vec{a}} )  ,
\end{align}
and hence our task is to calculate
\begin{align}
\frac{\partial^m}{\partial \beta^m}  \log \left[\frac{\tr_{V} (e^{-\beta H_{\vec{a}}})}{d_{V}}  \right]   &=- \tr_V \left(  H_{\vec{a}} \frac{\partial^{m-1}}{\partial \beta^{m-1}} \rho_{\vec{a}} \right).
\label{equiv_log_der1}
\end{align} 
By using Lemma~2 in Ref.~\cite{kuwahara2019gaussian}, we have 
\begin{align}
\frac{\partial^{m-1}}{\partial \beta^{m-1}} \tr \left(  H_{\vec{a}} \rho_{\vec{a}} \right)\Bigl|_{\beta=0} =\frac{(-1)^{m-1}}{d_{V}^{m}} 
\tr_{V^\co_{1:m}}\left( \tilde{H}_{\vec{a}}^{(0)}\tilde{H}_{\vec{a}}^{(1)}\cdots \tilde{H}_{\vec{a}}^{(m-1)}  \right),
\end{align} 
where in the inequality (B.3) in \cite{kuwahara2019gaussian}, we choose as $m_1=0$, $m_2=m-1$ and $\omega_\ed=H_{\vec{a}}$.
We thus obtain the equation~\eqref{in_the_case_of_L_co_eq_V}.
This completes the proof of Lemma~\ref{thm:express_Der_operator}. $\square$

 {~}

\hrulefill{\bf [ End of Proof of Proposition~\ref{prop:explicit_form_log_derivation} ] }

{~}

\begin{figure}
\centering
\includegraphics[clip, scale=0.6]{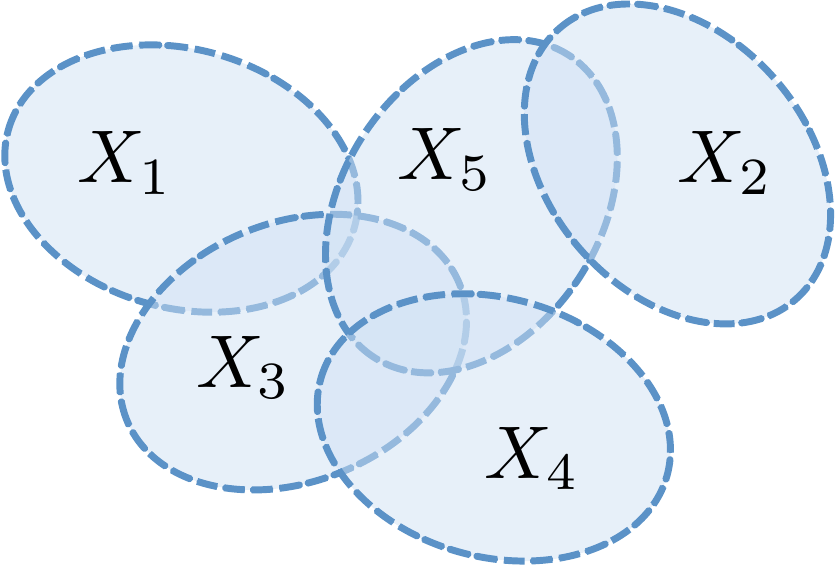}
\caption{$N_{\ed_s| w}$ is defined by a number of subsystems in $w$ that have overlap with $\ed_s$.
When $w=\{\ed_1,\ed_2,\ed_3,\ed_4,X_5\}$ is given as above, we have $N_{\ed_1 | w}=2$, $N_{\ed_2 | w}=1$, $N_{\ed_3 | w}=2$, $N_{\ed_4| w}=2$ and $N_{\ed_5| w}=4$.}
\label{fig:Number_overlap}
\end{figure}

We then aim to obtain an upper bound of $\left\| \tr_{L^\co _{1:m}} \left (\tilde{h}_{\ed_1}^{(0)} \tilde{h}_{\ed_2}^{(1)}\cdots \tilde{h}_{\ed_m}^{(m-1)} \right) \right\|$.
For the purpose, we utilize the following proposition.
\begin{prop}\label{prop:bound_on_Der}
Let $\{O_s\}_{s=0}^{m}$  be operators supported on a subset $w:=\{\ed_s\}_{s=0}^{m}$, respectively.
When they satisfy $\tr_{L^\co}(O_s)=0$ for $s=0,1,2\ldots,m$,
we obtain
\begin{align}
&\frac{1}{d_{L^\co}^{m}}\left \|\tr_{L^\co_{1:m}}  \left ( \tilde{O}_{0}^{(0)} \tilde{O}_{1}^{(1)} \tilde{O}_{2}^{(2)}\cdots \tilde{O}_{m-1}^{(m-1)}  \right) \right\| \le  \|O_{0}\|   \prod_{s=1}^{m} 
2 N_{\ed_s | w_L} \|O_{s}\| ,
\label{ineq_Der_norm_overlap_form}
\end{align}
where we define $\tilde{O}_s^{(s)}$ as in Eq.~\eqref{Def:tilde_O_s}.
$N_{\ed_s | w}$ is a number of subsets in $w$ that have overlap with $\ed_s$ (Fig.~\ref{fig:Number_overlap}):
\begin{align}
N_{\ed_s | w} = \# \{\ed \in w|  \ed\neq \ed_s ,  \ed \cap \ed_s \neq \emptyset  \}.
\end{align}
The proof is the same as that of Proposition~3 in Ref.~\cite{kuwahara2019gaussian}, which proves Ineq.~\eqref{ineq_Der_norm_overlap_form} for $L^\co=V$.
\end{prop}

In order to apply Proposition~\eqref{prop:bound_on_Der} to $\tr_{L^\co _{1:m}} \left (\tilde{h}_{\ed_1}^{(0)} \tilde{h}_{\ed_2}^{(1)}\cdots \tilde{h}_{\ed_m}^{(m-1)} \right)$,  the condition $\tr_{L^\co} (h_X)=0$ is necessary, whereas it is not generally satisfied.
Thus, instead of considering $h_X$, we consider $\mathfrak{h}_\ed$ which is defined as follows:
\begin{align}
\mathfrak{h}_\ed := h_\ed - \frac{h_\ed^L}{d_{L^\co}} \for X\in E_r,
\label{mathfrak_h_def}
\end{align}
where $\mathfrak{h}_\ed$ satisfies $\tr_{L^\co} (\mathfrak{h}_\ed) = \tr_{L^\co}(h_X) - h_\ed^L \tr_{L^\co} (\hat{1}) /d_{L^\co} =  h_\ed^L -h_\ed^L=0 $ from the definition~\eqref{definition:reduced_matrix}.
By using the notation of $\mathfrak{h}_\ed$, we obtain
\begin{align}
\tr_{L^\co_{1:m}} \left (\tilde{h}_{\ed_1}^{(0)} \tilde{h}_{\ed_2}^{(1)}\cdots \tilde{h}_{\ed_m}^{(m-1)} \right)
&= \tr_{L^\co_{1:m}} \left (\tilde{\mathfrak{h}}_{\ed_1}^{(0)} \tilde{\mathfrak{h}}_{\ed_2}^{(1)}\cdots \tilde{\mathfrak{h}}_{\ed_m}^{(m-1)} \right) 
+ \frac{h_{\ed_1}^L}{d_{L^\co}} \otimes \tr_{L^\co_{1:m}} \left (\tilde{\mathfrak{h}}_{\ed_2}^{(1)}\tilde{\mathfrak{h}}_{\ed_3}^{(2)}\cdots \tilde{\mathfrak{h}}_{\ed_m}^{(m-1)} \right),
\label{operator_average_0_trans}
\end{align}
where we use $\tilde{\mathfrak{h}}^{(s)}_\ed = \tilde{h}_\ed^{(s)}$ for $s\ge 1$ which comes from the definition~\eqref{Def:tilde_O_s}, and apply Eq.~\eqref{mathfrak_h_def} to $\tilde{h}_{\ed_1}^{(0)}$. 
We then prove $\tr_{L^\co_{1:m}} \left (\tilde{\mathfrak{h}}_{\ed_2}^{(1)}\tilde{\mathfrak{h}}_{\ed_3}^{(2)}\cdots \tilde{\mathfrak{h}}_{\ed_m}^{(m-1)} \right)=0$.  
By using the definition~\eqref{Def:tilde_O_s} for $\tilde{\mathfrak{h}}_{\ed_2}^{(1)}$, we have
\begin{align}
\tr_{L^\co_{1:m}} \left (\tilde{\mathfrak{h}}_{\ed_2}^{(1)}\tilde{\mathfrak{h}}_{\ed_3}^{(2)}\cdots \tilde{\mathfrak{h}}_{\ed_m}^{(m-1)} \right) 
= \tr_{L^\co_{1:m}} \left [ \left( \tilde{\mathfrak{h}}_{\ed_2,\tilde{\mathcal{H}}_1} -\tilde{\mathfrak{h}}_{\ed_2,\tilde{\mathcal{H}}_2} \right)\tilde{\mathfrak{h}}_{\ed_3}^{(2)} \cdots \tilde{\mathfrak{h}}_{\ed_m}^{(m-1)} \right] .
\label{operator_average_0_trans_second}
\end{align}
Because the operator $\tilde{\mathfrak{h}}_X^{(s)}$ ($s\ge 2$) is invariant under the swapping between the Hilbert spaces $H^{L^\co}_1$ and $H^{L^\co}_2$ (i.e., $\tilde{\mathcal{H}}_1 \leftrightarrow \tilde{\mathcal{H}}_2$), 
we have 
\begin{align}
\tr_{L^\co_{1:m}} \left( \tilde{\mathfrak{h}}_{\ed_2,\tilde{\mathcal{H}}_1}\tilde{\mathfrak{h}}_{\ed_3}^{(2)} \cdots \tilde{\mathfrak{h}}_{\ed_m}^{(m-1)} \right) 
= \tr_{L^\co_{1:m}} \left( \tilde{\mathfrak{h}}_{\ed_2,\tilde{\mathcal{H}}_2}\tilde{\mathfrak{h}}_{\ed_3}^{(2)} \cdots \tilde{\mathfrak{h}}_{\ed_m}^{(m-1)} \right) .
\end{align}
Therefore, the term~\eqref{operator_average_0_trans_second} vanishes and Eq.~\eqref{operator_average_0_trans} reduces to
\begin{align}
\tr_{L^\co_{1:m}} \left (\tilde{h}_{\ed_1}^{(0)} \tilde{h}_{\ed_2}^{(1)}\cdots \tilde{h}_{\ed_m}^{(m-1)} \right)
&= \tr_{L^\co_{1:m}} \left (\tilde{\mathfrak{h}}_{\ed_1}^{(0)} \tilde{\mathfrak{h}}_{\ed_2}^{(1)}\cdots \tilde{\mathfrak{h}}_{\ed_m}^{(m-1)} \right) .
\label{operator_average_0_trans_equal}
\end{align}

By using Proposition~\ref{prop:bound_on_Der}, we obtain an upper bound of $\tr_{L^\co_{1:m}} \left (\tilde{h}_{\ed_1}^{(0)} \tilde{h}_{\ed_2}^{(1)}\cdots \tilde{h}_{\ed_m}^{(m-1)} \right)$ as follows:
\begin{align}
\frac{1}{d_{L^\co}^{m}} \left\| \tr_{L^\co_{1:m}} \left (\tilde{h}_{\ed_1}^{(0)} \tilde{h}_{\ed_2}^{(1)}\cdots \tilde{h}_{\ed_m}^{(m-1)} \right)\right \|
=&\frac{1}{d_{L^\co}^{m}} \left\|  \tr_{L^\co_{1:m}} \left (\tilde{\mathfrak{h}}_{\ed_1}^{(0)} \tilde{\mathfrak{h}}_{\ed_2}^{(1)}\cdots \tilde{\mathfrak{h}}_{\ed_m}^{(m-1)} \right) \right \| \notag \\
\le & \|\mathfrak{h}_{\ed_1}\|  \prod_{s=2}^{m} 2  N_{\ed_{s} | w} \|\mathfrak{h}_{\ed_s}\| 
\le \frac{1}{2}\prod_{s=1}^{m} 4 N_{\ed_{s} | w} \|h_{\ed_s}\| , \label{bound_operator_product_extended_space}
 \end{align}
where we use $\| \mathfrak{h}_\ed\| \le 2 \|h_\ed\|$ which comes from the definition~\eqref{mathfrak_h_def}.
By combining the inequality~\eqref{bound_operator_product_extended_space} with Eq.~\eqref{simple_expression_der_w_log}, we obtain an upper bound of
\begin{align}
\left \|\Der_{w}\log \tilde{\rho}^{L}_{\vec{a}} \bigl|_{\vec{a}=\vec{0}} \right \| 
\le \frac{1}{2} \prod_{s=1}^{m} 4\beta  N_{\ed_{s} | w} \|h_{\ed_s}\| .\label{eq: log_bound_der_prop}
\end{align}
By applying the inequality~\eqref{eq: log_bound_der_prop} to the cases $L=AB$, $L=BC$, $L=ABC$ and $L=B$, 
we obtain the following inequality:
\begin{align}
\left\| \Der_{w}  \tilde{H}_{\vec{a}}(A:C|B)  \bigl|_{\vec{a}=\vec{0}} \right\| \le  2 (4\beta)^m   \prod_{s=1}^{m}  N_{\ed_{s} | w} \|h_{\ed_s}\|, \label{eq: log_bound_der_mutual_info}
 \end{align}
 where $\tilde{H}_{\vec{a}}(A:C|B)$ has been defined in Eq.~\eqref{def_tilde_H_a_A_C_B}.
Then, the final task is to upper-bound the summation with respect to $\sum_{w \in \Gc^{A,C}_{r,m}}$ in Eq.~\eqref{generalized_cluster_exp_tilde_H}:
\begin{align}
\left\|  \tilde{H}_{\vec{1}}(A:C|B)\right\| & \le \sum_{m=1}^{\infty}\frac{1}{m!} \sum_{w \in \Gc^{A,C}_{r,m}} n_w \left\| \Der_{w}  \tilde{H}_{\vec{a}}(A:C|B)  \Bigl|_{\vec{a}=\vec{0}} \right\| \notag \\
&\le \sum_{m=1}^{\infty}\frac{2 (4\beta)^m }{m!} \sum_{w \in \Gc^{A,C}_{r,m}} n_w  \prod_{s=1}^{m}  N_{\ed_{s} | w} \|h_{\ed_s}\|  , 
\label{upper_bound_mutual_info_last}
\end{align}
where we use the proposition~\ref{prop: connceted_cluster_mutual} in the first inequality.

For the estimation of the summation, we first focus on the fact that 
any cluster in $w \in \Gc^{A,C}_{r,m}$ must have overlaps with the surface regions of $A$ and $C$, say $\partial A_r$ and $\partial C_r$ ($r\in \mathbb{N}$):
\begin{align}
\partial A_r := \{v\in A| \dist_{v,A^\co}\le r \},\quad \partial C_r := \{v\in C | \dist_{v,C^\co} \le r \}.
\label{def_surface_region_supp}
\end{align}
Second, because $ \dist_{A,C}$ is the minimum path length on the graph $(V,E)$ to connect the subsystems $A$ and $C$, 
the condition $w \in \Gc^{A,C}_{r,m}$ implies $|w| \ge \dist_{A,C}/r$ as the necessary condition.
From these two fact, we will replace the summation $\sum_{w \in \Gc^{A,C}_{r,m}}$ with $\sum_{v \in \partial A_r} \sum_{m \ge \dist_{A,C}/r} \sum_{w \in \Gc^{v}_{r,m}}$ by taking all the clusters with the sizes $|w| \ge \dist_{A,C}/r$ which have overlap with $A$ into account:
\begin{align}
\sum_{m=1}^{\infty}\frac{2 (4\beta)^m }{m!} \sum_{w \in \Gc^{A,C}_{r,m}} n_w  \prod_{s=1}^{m}  N_{\ed_{s} | w} \|h_{\ed_s}\|  
\le \sum_{v \in \partial A_r} \sum_{m \ge \dist_{A,C}/r} 2 (4\beta)^m   \sum_{w \in \Gc^{v}_{r,m}}\frac{n_w}{m!} \prod_{s=1}^{m}  N_{\ed_{s} | w} \|h_{\ed_s}\|     , \label{mutual_sum_decomp_log}
 \end{align}
where the same inequality holds for the replacement of $\sum_{v \in \partial A_r}$ by $\sum_{v \in \partial C_r}$.

In order to estimate the summation of $\sum_{w \in \Gc^{v}_{r,m}}$, we utilize the following proposition which has been given in Ref.~\cite{kuwahara2019gaussian}:
\begin{prop}[Proposition~4 in Ref.~\cite{kuwahara2019gaussian}] \label{prop: convergence of cluster expansion}
Let $\{o_\ed\}_{\ed\in E_\infty}$ be arbitrary operators such that 
\begin{align}
\sum_{\ed| \ed \ni v}\| o_{\ed}\|\le g     \for \forall v\in V ,\label{cond_for_o_ed}
\end{align}
where $E_\infty$ is defined by Eq.~\eqref{eq:ham_graph_sup2} and it gives the set of all the subsystems $X\subset V$ with $|X| \le k$. 
Then, for an arbitrary subset $L$, we obtain
\begin{align}
\sum_{w \in \Gc^L_{m}} \frac{n_{w} }{m!} \prod_{s=1}^{m} N_{\ed_s | w_L}  \| o_{\ed_{s}}\|    \le \frac{1}{2}  e^{|L|/k} (2e^{3} gk)^m ,
\label{Proposition_4 in Ref.kuwahara2019gaussian}
 \end{align}
 where $w_L$ is defined as $w_L : = \{L,\ed_1,\ed_2 , \ldots, \ed_{|w|}\}$ for $w = \{\ed_1,\ed_2 , \ldots, \ed_{|w|}\}$.
\end{prop}

By applying Proposition~\ref{prop: convergence of cluster expansion} to the inequality~\eqref{mutual_sum_decomp_log}, we have
\begin{align}
\sum_{w \in \Gc^{v}_{r,m}}    \frac{n_w}{m!}   \prod_{s=1}^{m}  N_{\ed_{s} | w} \|h_{\ed_s}\|   \le \frac{1}{2}  e^{1/k} (2e^{3} k)^m, \label{Prop:cluster_summation_region_L1}
 \end{align}
where we use $N_{\ed_s | w_L} \le N_{\ed_s | w}$ in \eqref{Proposition_4 in Ref.kuwahara2019gaussian} and the condition~\eqref{eq:ham_graph_sup} gives $g=1$.
Therefore, the inequality~\eqref{mutual_sum_decomp_log} reduces to
\begin{align}
\sum_{m=0}^{\infty}\frac{1}{m!} \sum_{w \in \Gc^{A,C}_{r,m}} n_w\left\| \Der_{w}   \tilde{H}_{\vec{a}}(A:C|B)  \bigl\|_{\vec{a}=\vec{0}} \right| 
&\le  \sum_{v \in \partial A_r} \sum_{m \ge \dist_{A,C}/r} e^{1/k} (8e^{3} k\beta)^m   \notag \\
&\le e |\partial A_r| \frac{(8e^{3} k\beta)^{\dist_{A,C}/r}}{1-8e^{3} k\beta}    , \label{summation_w_v_mutual_cond}
\end{align}
where we use $k\ge 1$.
We notice that the same inequality holds for the replacement of $|\partial A_r| $ by $|\partial C_r|$.
 By combining the inequalities~\eqref{cond_mutual_info_tilde_rho_upper_bound}, \eqref{upper_bound_mutual_info_last} and \eqref{summation_w_v_mutual_cond}, we prove Theorem~\ref{main_theorem_QAMC_mutual_information}.
$\square$

\section{Quasi-Locality of effective Hamiltonian on a subsystem: Proof of Theorem~\ref{main_lemma_effective_Ham}}

We here consider the effective Hamiltonian on a subsystem $L$, which we define as
\begin{align}
\tilde{H}_L := -\beta^{-1} \log \tilde{\rho}^{L} , 
\end{align}
where $\tilde{\rho}^{L}$ is defined in Eq.~\eqref{def_rho_L_tilde_rho_L}.
We prove the following theorem which refines the Theorem~\ref{main_lemma_effective_Ham}: 
\begin{theorem} \label{thm:locality_exp_effectiveHam}
The effective Hamiltonian $\tilde{H}_{L} $ is given by a quasi-local operator
 \begin{align}
\tilde{H}_L  =H_L  + \sum_{m=1}^{\infty}\sum_{w \in \Gc^{L,L^\co}_{r,m}} n_w h_{L_w} - \frac{\hat{1}}{\beta} \log {Z_{L^\co}} 
\label{effective_Hamiltonian_H_L_theorem_quasi-local}
\end{align}
with 
 \begin{align}
H_L:=\sum_{\ed\subset L} h_\ed  , \quad   Z_{L^\co}:=\frac{1}{d_L}\tr  (e^{-\beta H_{L^\co}} \otimes \hat{1}_{L}) 
\end{align}
for $L \subset V$,
where each of $\{h_{L_w}\}_{w \in \Gc^{L,L^\co}_{r,m}}$ is supported on the subsystem $L_w:= L\cap V_w$ (see Def.~\eqref{def:h_w_local}) and 
$\Gc^{L,L^\co}_{r,m}$ is defined as a cluster subset defined in Def.~\ref{Def:Connected_cluster_to_A_B}. 
The effective interaction terms $\{h_{L_w}\}_{w \in \Gc^{L,L^\co}_{r,m}}$ is exponentially localized around the boundary:
 \begin{align}
\sum_{m>m_0}^{\infty} \sum_{w \in \Gc^{L,L^\co}_{r,m}} n_w\| h_{L_w}\| 
&\le  \frac{e}{4\beta}  \frac{(\beta/\beta_c)^{m_0+1}}{1- \beta/\beta_c} |\partial L_r|   \label{norm_sum_effective_ham}
\end{align}
for an arbitrary $m_0$.
\end{theorem}

From Eq.~\eqref{effective_Hamiltonian_H_L_theorem_quasi-local}, the effective interaction term $\Phi_L$ is given by
 \begin{align}
\Phi_L =\sum_{m=1}^{\infty}\sum_{w \in \Gc^{L,L^\co}_{r,m}} n_w h_{L_w} - \frac{\hat{1}}{\beta} \log {Z_{L^\co}}  .
\end{align}
Because of $\diam ( V_w) \le mr$, the subsystem $L\cap V_w$ ($w \in \Gc^{L,L^\co}_{r,m}$) is separated from the boundary $\partial L$ at most by a distance $mr$, 
namely $L\cap V_w \subseteq \partial L_{mr}$, where $\partial L_{l}$ has been defined in Eq.~\eqref{surface_region_of_L} as follows:
\begin{align}
\partial L_l :=\{v\in L| \dist_{v,L^\co}\le l \} . \label{surface_region_of_L_supp}
\end{align}
Hence, by defining $\Phi_{\partial L_l}$ as 
 \begin{align}
\Phi_{\partial L_l} =  \sum_{m\le \lfloor l/r \rfloor} \sum_{w \in \Gc^{L,L^\co}_{r,m}} n_w h_{L_w} - \frac{\hat{1}}{\beta} \log {Z_{L^\co}}  ,
\end{align}
we have 
 \begin{align}
\| \Phi_L  -\Phi_{\partial L_l}\| \le  \frac{e}{4\beta}  \frac{(\beta/\beta_c)^{l/r}}{1- \beta/\beta_c} |\partial L_r| .
\end{align}
This gives the proof of Theorem~\ref{main_lemma_effective_Ham}.

\subsection{Proof of Theorem~\ref{thm:locality_exp_effectiveHam}}

In order to apply the generalized cluster expansion, we first parametrize $\tilde{H}_L$ as 
\begin{align}
\tilde{H}_{L,\vec{a}} :=  -\beta^{-1}  \log \tilde{\rho}^{L}_{\vec{a}}. 
\end{align}
As in Eq.~\eqref{Cluster_decomp_mutual}, the generalized cluster expansion for $\tilde{H}_{L,\vec{a}}$ reads 
\begin{align}
\tilde{H}_{L,\vec{1}}  =- \frac{1}{\beta}\sum_{m=0}^{\infty}\frac{1}{m!} \sum_{w \in \Cl_{r,m}} n_w \Der_{w} \tilde{H}_{L,\vec{a}} \Bigl|_{\vec{a}=\vec{0}}. \label{Cluster_decomp_mutual_effectiveHam}
\end{align}
We can now prove the following proposition:
\begin{prop} \label{prop:generalized_cluster_exp_effectiveHam}
The summation with respect to the clusters $\sum_{w \in \Cl_{r,m}}$ reduces to  the following form: 
 \begin{align}
\tilde{H}_{L,\vec{1}}  =H_L  - \frac{1}{\beta}\log Z_{L^\co} +\sum_{m=1}^{\infty}\frac{1}{m!} \sum_{w \in \Gc^{L,L^\co}_{r,m}} n_w \Der_{w} \tilde{H}_{L,\vec{a}} \Bigl|_{\vec{a}=\vec{0}}  ,\label{Cluster_decomp_effectiveHam_connected_one}
\end{align}
where $H_L:=\sum_{\ed\subset L} h_\ed$ and $Z_{L_\co}:=d_L^{-1} \tr (e^{-\beta H_{L^\co}})$.
\end{prop}

\subsubsection{Proof of Proposition~\ref{prop:generalized_cluster_exp_effectiveHam}} \label{proof_connected_cluster_contribute}

For the proof, 
we first prove 
\begin{align}
\Der_{w}  \log(\tilde{\rho}_{\vec{a}_w}^{L}) =0 \for w \notin \Gc_{r,|w|}. \label{connected_cluster_vanish_effectiveham}
\end{align}
The proof is given as follows.
Due to the existence of decomposition $w=w_1 \sqcup w_2$ such that $V_{w_1}\cap V_{w_2} =\emptyset$, we have
$e^{-\beta H_{\vec{a}_w}} = e^{-\beta H_{\vec{a}_{w_1}}}  \otimes e^{-\beta H_{\vec{a}_{w_2}}}$ and hence, 
\begin{align}
\log (\tilde{\rho}_{\vec{a}_w}^{L} ) = \log (\tilde{\rho}_{\vec{a}_{w_1}}^{L} ) + \log (\tilde{\rho}_{\vec{a}_{w_2}}^{L} ) - \log  d_{L^\co}  .
\end{align}
Because $\Der_{w_2}\log (\tilde{\rho}_{\vec{a}_{w_1}}^{L} ) = \Der_{w_1}\log (\tilde{\rho}_{\vec{a}_{w_2}}^{L})=0$, we obtain Eq.~\eqref{connected_cluster_vanish_effectiveham}.

We then consider the cases of $V_w\subseteq L$ and $V_w \subseteq L^\co$ in Eq.~\eqref{Cluster_decomp_mutual_effectiveHam}. 
In the case of $V_w\subseteq L$, the definition~\eqref{def_rho_L_tilde_rho_L} gives 
\begin{align}
\log(\tilde{\rho}_{L,\vec{a}_w}) = -\beta H_{\vec{a}_w} + \log  d_{L^\co}.
\end{align}
Therefore, we have $\Der_w \log(\tilde{\rho}_{\vec{a}_w}^{L})$ vanishes for $m\ge 2$, and 
\begin{align}
- \frac{1}{\beta}\sum_{m=1}^{\infty}\frac{1}{m!} \sum_{w \in \Gc_{r,m}, V_w\subseteq L} n_w \Der_{w} \tilde{H}_{L,\vec{a}} \Bigl|_{\vec{a}=\vec{0}} = \sum_{\ed \subset L} h_{\ed}= H_L .
\label{der_H_L_prop}
\end{align}
On the other hand, in the case of $V_w\subseteq L^\co$, $\log(\tilde{\rho}_{\vec{a}_w}^{L})$ becomes a constant operator (i.e., $\log(\tilde{\rho}_{\vec{a}_w}^{L})\propto\hat{1}$). 
Hence, we obtain
\begin{align}
- \frac{1}{\beta} \sum_{m=0}^{\infty}\frac{1}{m!} \sum_{w \in \Gc_{r,m}, V_w\subseteq L^\co} n_w \Der_{w} \tilde{H}_{L,\vec{a}} \Bigl|_{\vec{a}=\vec{0}} 
&=- \frac{1}{\beta} \log  [\tr_{L^\co} (e^{ -\beta H_{L^\co}})] = - \frac{\log Z_{L_\co}}{\beta}  \hat{1}.
\end{align}

Thus, the summation~\eqref{Cluster_decomp_mutual_effectiveHam} reduces to
\begin{align}
\tilde{H}_{L,\vec{1}}  =&- \frac{1}{\beta}\sum_{m=1}^{\infty}\frac{1}{m!} \sum_{w \in \Gc_{r,m}, V_w\subseteq L} n_w \Der_{w} \tilde{H}_{L,\vec{a}} \Bigl|_{\vec{a}=\vec{0}} 
- \frac{1}{\beta} \sum_{m=0}^{\infty}\frac{1}{m!} \sum_{w \in \Gc_{r,m}, V_w\subseteq L^\co} n_w \Der_{w} \tilde{H}_{L,\vec{a}} \Bigl|_{\vec{a}=\vec{0}} \notag \\
&- \frac{1}{\beta} \sum_{m=1}^{\infty}\frac{1}{m!} \sum_{w \in \Gc^{L,L^\co}_{r,m}} n_w \Der_{w} \tilde{H}_{L,\vec{a}} \Bigl|_{\vec{a}=\vec{0}} \notag \\
=&H_L  - \frac{1}{\beta} \log Z_{L^\co} - \frac{1}{\beta} \sum_{m=1}^{\infty}\frac{1}{m!} \sum_{w \in \Gc^{L,L^\co}_{r,m}} n_w \Der_{w} \tilde{H}_{L,\vec{a}} \Bigl|_{\vec{a}=\vec{0}} .
\label{expansion_cluster_gene_eff}
\end{align}
This completes the proof. $\square$

 {~}

\hrulefill{\bf [ End of Proof of Proposition~\ref{prop:generalized_cluster_exp_effectiveHam} ] }

{~}

We now define $h_{L_w}$ as 
 \begin{align}
h_{L_w}: =\frac{-\beta^{-1}}{m!}   \Der_{w} \tilde{H}_{L,\vec{a}} \Bigl|_{\vec{a}=\vec{0}}, \label{def:h_w_local}
\end{align}
where $w \in \Gc^{L,L^\co}_{r,m}$. Note that the operator $h_{L_w}$ is supported on the subsystem $L_w= L \cap V_w$. 
Then, the effective Hamiltonian $\tilde{H}_{L,\vec{1}} $ is formally written by 
 \begin{align}
\tilde{H}_{L,\vec{1}}  =H_L  - \frac{1}{\beta} \log Z_{L^\co} +\sum_{m=1}^{\infty} \sum_{w \in \Gc^{L,L^\co}_{r,m}} n_wh_{L_w}.
\end{align}

By using the proposition~\ref{prop:explicit_form_log_derivation} with the inequalities~\eqref{bound_operator_product_extended_space} and \eqref{Prop:cluster_summation_region_L1}, 
we have 
 \begin{align}
\sum_{w \in \Gc^{v}_{r,m}}n_w \| h_{L_w}\|
&\le  \frac{\beta^{-1}}{m!} \sum_{w \in \Gc^{v}_{r,m}} \frac{n_w}{2} \prod_{s=1}^{m} 4 \beta N_{\ed_{s} | w} \|h_{\ed_s}\|  \notag \\
&\le  (4 \beta)^{m-1} e^{1/k}(2e^{3} k)^{m}\le  \frac{e}{4\beta}  (\beta/\beta_c)^{m} ,
\end{align}
where we use $e^{1/k}\le e$ due to $k\ge 1$.
By using the above inequality, the contribution of $m$th order terms in the expansion~\eqref{Cluster_decomp_effectiveHam_connected_one} is bounded from above by
 \begin{align}
\sum_{w \in \Gc^{L,L^\co}_{r,m}} n_w\| h_{L_w}\| 
&\le \sum_{v\in \partial{L}_r} \sum_{w \in \Gc^{v}_{r,m}}  n_w\| h_{L_w}\| \le \frac{e}{4\beta}  (\beta/\beta_c)^{m} |\partial L_r| , \label{est_1_eff}
\end{align}
where $\partial{L}_r$ has been defined in Eq.~\eqref{surface_region_of_L_supp}.
 \begin{align}
\sum_{m>m_0}^{\infty} \sum_{w \in \Gc^{L,L^\co}_{r,m}} n_w\| h_{L_w}\| 
&\le   \frac{e  |\partial L_r|}{4\beta} \sum_{m=m_0+1}^{\infty} (\beta/\beta_c)^{m} = \frac{e  |\partial L_r|}{4\beta}  \frac{(\beta/\beta_c)^{m_0+1}}{1-\beta/\beta_c}.
\label{est_2_eff}
\end{align}
This completes the proof of Theorem~\ref{thm:locality_exp_effectiveHam}.
$\square$


%

\subsection{Computational cost of cluster summation}

We here show the computational cost to estimate the effective Hamiltonian $\tilde{H}_{L}$.
For this aim, we start from a slightly weaker expression than Eq.~\eqref{Cluster_decomp_effectiveHam_connected_one} as follows
 \begin{align}
\Phi_L= \tilde{H}_{L,\vec{1}} -H_L = - \frac{1}{\beta} \sum_{m=0}^{\infty}\frac{1}{m!} \sum_{w \in \Gc_{r,m}, V_w\subseteq L^\co} n_w \Der_{w} \tilde{H}_{L,\vec{a}} \Bigl|_{\vec{a}=\vec{0}} - \frac{1}{\beta} \sum_{m=1}^{\infty}\frac{1}{m!} \sum_{w \in \Gc^{L,L^\co}_{r,m}} n_w \Der_{w} \tilde{H}_{L,\vec{a}} \Bigl|_{\vec{a}=\vec{0}}  ,\label{Cluster_decomp_effectiveHam_connected_comp}
\end{align}
where we use the second and third terms in the first equation of~\eqref{expansion_cluster_gene_eff}.
Our task is to estimate the computational cost of 
$
n_w \Der_{w} \tilde{H}_{L,\vec{a}} \bigl|_{\vec{a}=\vec{0}} 
$
and the number of clusters in $\{ w \in \Gc_{r,m}, V_w\subseteq L^\co\}$ and $w \in \Gc^{L,L^\co}_{r,m}$.

First, we consider $n_w \Der_{w} \tilde{H}_{L,\vec{a}} \bigl|_{\vec{a}=\vec{0}}$.
As defined in Eq.~\eqref{Cluster_decomp_mutual}, $n_w$ is immediately calculated, and hence we need to estimate the computational cost to calculate the multiderivative
\begin{align}
\Der_{w} \tilde{H}_{L,\vec{a}_{w}} \Bigl|_{\vec{a}_w=\vec{0}} = \prod_{j=1}^m \frac{\partial}{\partial a_{\ed_j}}\tilde{H}_{L,\vec{a}_{w}} \Bigl|_{\vec{a}_w=\vec{0}}
\end{align}
with $w=\{\ed_s\}_{s=1}^m$ by using numerical differentiation.
The operator $\tilde{H}_{L,\vec{a}_w}$ is given by
\begin{align}
\tilde{H}_{L,\vec{a}_w} = -\beta^{-1}  \log \tilde{\rho}^{L}_{\vec{a}_w} =  -\beta^{-1}  \tr_{L^\co} \left( e^{-\beta H_{\vec{a}_w}} \right) \otimes \hat{1}_{L^\co} ,
\end{align}
where we use the definition~\eqref{def_rho_L_tilde_rho_L}.
Note that $H_{\vec{a}_{w}}$ is supported on $V_w \subset V$. Hence, the computational cost to calculate $\tilde{H}_{L,\vec{a}_w}$ is at most of $d^{\orderof{|V_w|}}$. 
In order to perform the differentiation, we need to calculate $2^{|w|}$ values of $\tilde{H}_{L,\vec{a}_w} $ for $a_{\ed_s} = \pm \Delta$ ($\Delta \to +0$) for $s=1,2, \ldots, |w|$.
Thus, for the numerical differentiation we need the computational cost of $2^{|w|} \cdot d^{\orderof{|V_w|}}= d^{\orderof{mk}}$ with $|w|=m$, where we use $|V_w|\le |w| k$.

We then need to sum up the contributions from all the clusters in $\{ w \in \Gc_{r,m}, V_w\subseteq L^\co\}$ and $w \in \Gc^{L,L^\co}_{r,m}$.
For the purpose, we first prove the following theorem on the number of clusters:
\begin{prop} \label{prop:cluster_counting}
The total number of different clusters in $\Gc^{L^\co}_{r,m}$ is bounded as follows: 
 \begin{align}
\# \left\{w\in \Cl_{r,m} \bigl | w \in \Gc_{r,m}, V_w\subseteq L^\co \quad {\rm or}\quad  w \in \Gc^{L,L^\co}_{r,m} \right\}  \le |L^\co|\left(3 \cdot 2^k d_G^{rk}\right)^{m} .
\label{ineq:cluster_counting}
 \end{align}
This roughly gives the total number by $|L^\co| d_G^{\orderof{rkm}}$,  
\end{prop}

In total, the computation of the $m$-th order in the expansion~\eqref{Cluster_decomp_effectiveHam_connected_comp} is performed with the runtime bounded from above by
\begin{align}
d^{\orderof{mk}} \cdot  |L^\co| d_G^{\orderof{rkm}} \le n (d\cdot d_G^r)^{mk}.
\end{align}
Also, the convergence of the expansion~\eqref{Cluster_decomp_effectiveHam_connected_comp} is estimated as in~\eqref{est_1_eff} and \eqref{est_2_eff}
 \begin{align}
&\sum_{w \in \Gc_{r,m}, V_w\subseteq L^\co} \left \|n_w \Der_{w} \tilde{H}_{L,\vec{a}} \Bigl|_{\vec{a}=\vec{0}}\right\| -  \sum_{m=1}^{\infty}\frac{1}{m!} \sum_{w \in \Gc^{L,L^\co}_{r,m}} \left\| n_w \Der_{w} \tilde{H}_{L,\vec{a}} \Bigl|_{\vec{a}=\vec{0}} \right\|  \notag \\
\le & \sum_{v\in L^\co} \sum_{w \in \Gc^{v}_{r,m}}  n_w\| h_{L_w}\| \le  \frac{e}{4\beta}  (\beta/\beta_c)^{m}|L^\co| \le \frac{e}{4\beta}  (\beta/\beta_c)^{m}n,
\end{align}
which yields 
 \begin{align}
\sum_{m>m_0}^{\infty} \sum_{w \in \Gc_{r,m}, V_w\subseteq L^\co} \left \|n_w \Der_{w} \tilde{H}_{L,\vec{a}} \Bigl|_{\vec{a}=\vec{0}}\right\| - \sum_{m>m_0}^{\infty}  \sum_{m=1}^{\infty}\frac{1}{m!} \sum_{w \in \Gc^{L,L^\co}_{r,m}} \left\| n_w \Der_{w} \tilde{H}_{L,\vec{a}} \Bigl|_{\vec{a}=\vec{0}} \right\| 
&\le    \frac{en}{4\beta}  \frac{(\beta/\beta_c)^{m_0+1}}{1-\beta/\beta_c}.
\end{align}

Therefore, we need to choose $m=\orderof{\log (1/\epsilon)}$ to calculate $\Phi_L$ up to an error $n\epsilon$ as long as $\beta<\beta_c$. 
Hence, the computational cost is estimated as
\begin{align}
n (d\cdot d_G^r)^{k\orderof{\log (1/\epsilon)}} = n (1/\epsilon)^{\orderof{k \log (d d_G^r )}}.
\end{align}
This completes the derivation of the computational cost~\eqref{computational_cost_main} for computing $\Phi_L$. $\square$

\subsubsection{Proof of Proposition~\ref{prop:cluster_counting}}

We here prove Proposition~\ref{prop:cluster_counting} which gives an upper bound of the number of cluster connecting to a subset $L^\co$.
For the purpose, we estimate the number of clusters in $\Gc^{v}_{r,m}$, which gives an upper bound of 
 \begin{align}
\# \left\{w\in \Cl_{r,m} \bigl | w \in \Gc_{r,m}, V_w\subseteq L^\co \quad {\rm or}\quad  w \in \Gc^{L,L^\co}_{r,m} \right\} \le  \sum_{v\in L^\co} \# \left\{w \bigl | w \in \Gc^{v}_{r,m}  \right\}  .
\label{start_upp_counting}
 \end{align}

First, we count the number of clusters $w=\{\ed_s\}_{s=1}^q$ which satisfy $\ed_s \cap Y \neq \emptyset$ for $\forall \ed_s $ ($s=1,2,\ldots, q$), where $Y$ is an arbitrary subset in $V$.
The number is bounded from above by
\begin{align}
\#\left\{ w\in \Cl_{r,q} | \ed_s \cap Y \neq \emptyset ,\ s=1,2,\ldots,q \right\}\le \sum_{\{ v_1,v_2,\ldots,v_q\} \subseteq Y} \prod_{s=1}^q {\rm deg}(v_s) , \label{Num_upper_bound_cluster_connect_L}
\end{align}
where we define ${\rm deg}(v)$ as ${\rm deg}(v):=\# \left\{ \ed\in E_r |\ed \ni v\right\}$. 
By using the graph degree $d_G$, we can upper-bound ${\rm deg}(v)$ by
\begin{align}
{\rm deg}(v)=\# \left\{ \ed\in E_r |\ed \ni v\right\} \le \binom{d_G^r}{k}\le d_G^{rk}, \label{deg_V_ineq}
\end{align}
where $d_G^r$ is the upper bound of the number of vertices $\{v'\}_{v'\in V}$ such that $\dist_{v,v'} \le r$.
Also, note that $\ed\in E_r$ implies $|\ed| \le k$ from the definitions~\eqref{eq:ham_graph_sup} and \eqref{eq:ham_graph_sup2}.
The summation with respect to $\{ v_1,v_2,\ldots , v_{q} \}$ is equal to the $q_1$-multicombination from a set of $|L|$ vertices, which is equal to
\begin{align} 
\sum_{\{ v_1,v_2,\ldots,v_q\} \subseteq Y} =\multiset{|Y|}{q} = \binom{q+|Y|-1}{q} \le  2^{q+|Y|-1}\label{comb_q_1_q_2__q_l_count}.
\end{align}
By combining the inequalities~\eqref{deg_V_ineq} and \eqref{comb_q_1_q_2__q_l_count} with \eqref{Num_upper_bound_cluster_connect_L}, we obtain 
\begin{align}
\#\left\{ w\in \Cl_{r,q}  | \ed_s \cap Y \neq \emptyset ,\ s=1,2,\ldots,q \right\}  \le  2^{|Y|-1} (2d_G^{rk})^{q}.
\end{align}

\begin{figure}
\centering
\includegraphics[clip, scale=0.6]{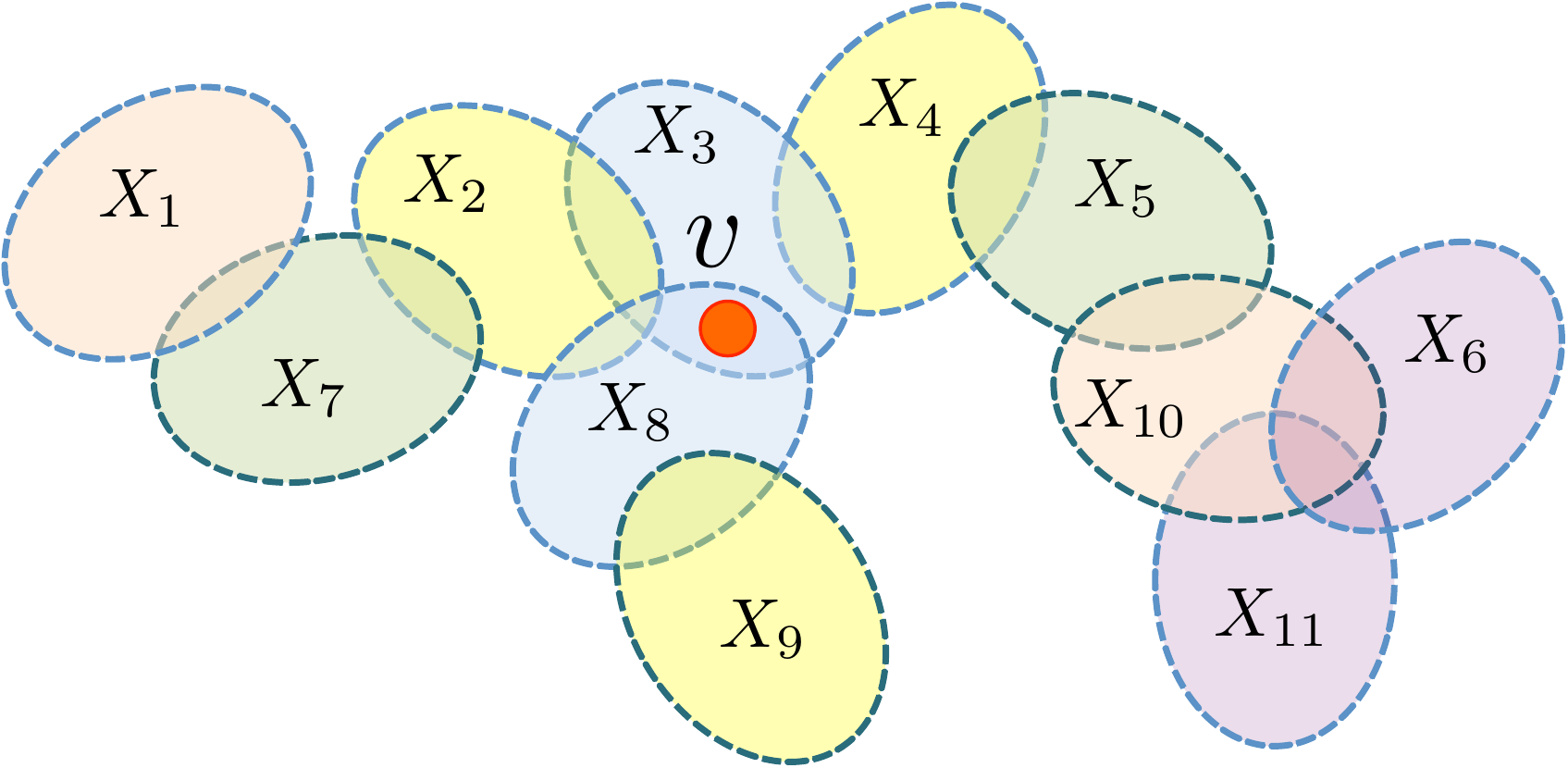}
\caption{Decomposition of $w$ in $\Gc^{v}_{r,m}$ as in Eq.~\eqref{w_L_decomposition_cal_cluster_sum}.
In the picture, we have $w_0=\{\ed_3,\ed_8\}$, $w_1=\{\ed_2,\ed_4,\ed_9\}$, $w_2=\{\ed_5,\ed_7\}$, $w_3=\{\ed_1,\ed_{10}\}$, $w_4=\{\ed_6,\ed_{11}\}$.
}
\label{fig:connection_hierarchy}
\end{figure}

We then consider the following decomposition of  $w \in \Gc^{v}_{r,m}$ (see Fig.~\ref{fig:connection_hierarchy}):
\begin{align}
w= w_0 \sqcup w_1 \sqcup w_2 \sqcup \cdots \sqcup w_l, \quad 0\le l\le m-1,\label{w_L_decomposition_cal_cluster_sum}
\end{align}
where $w_j \subset w_L $ satisfy $\dist(w_j, v) = j$ for $j=0,1,2,\ldots,l$. 
Here, we define $\dist(w_j, w_0)$ as the shortest path length in the cluster $w_0\sqcup w_1 \sqcup \cdots \sqcup w_{j-1}$ which connects from $w_j$ to $v$. 
We also define $q_j:=|w_j|$ with $q_j\ge 1$.
We notice that all the clusters $w \in \Gc^{v}_{r,m}$ can be decomposed into the from of \eqref{w_L_decomposition_cal_cluster_sum}.

For fixed $\{q_0,q_1,\ldots,q_l\}$, the number of clusters $\{w_1,w_2,\ldots,w_l\}$ defined as in Eq.~\eqref{w_L_decomposition_cal_cluster_sum} is bounded by
\begin{align}
&\#\left\{ w\in \Cl_{r,q_0}  | \ed_{0,s} \cap v \neq \emptyset ,\ s=1,2,\ldots,q_0 \right\}  \prod_{j=1}^l \max_{w_{j-1} \in \Cl_{r,q_{j-1}}} \left(\#\left\{ w\in \Cl_{r,q_j}  | \ed_{s_j} \cap V_{w_{j-1}} \neq \emptyset ,\ s_j=1,2,\ldots,q_j \right\}  \right) \notag \\
\le&  (2d_G^{rk})^{q_0} \prod_{j=1}^l \left[ 2^{kq_{j-1}-1} (2d_G^{rk})^{q_j}\right] \le 2^{-l}  \left(2^{k+1}d_G^{rk}\right)^{m},
\end{align}
where we denote $w_j=\{\ed_{s_j} \}_{s_j=1}^{q_j}$; note that $\sum_{j=0}^l q_j =m$.
Then, by taking the summation with respect to $\{q_0,q_1,\ldots,q_l\}$ and $l$, we finally obtain the upper bound of $\#\{w| w\in \Gc_{r,m}^v\} $ as follows:
\begin{align}
\#\{w| w\in \Gc_{r,m}^v\} &\le \sum_{l=0}^{m-1} \sum_{\substack{q_0+q_1+\cdots+q_l=m\\q_0\ge1, q_1\ge 1,\ldots, q_l\ge 1}}2^{-l} \left(2^{k+1}d_G^{rk}\right)^{m}\notag \\
&= \sum_{l=0}^{m-1} \multiset{l+1}{m-l-1} 2^{-l}\left(2^{k+1}d_G^{rk}\right)^{m}  \notag \\
&=\sum_{l=0}^{m-1} \binom{m-1}{l} 2^{-l} \left(2^{k+1}d_G^{rk}\right)^{m}  \le  \left(3 \cdot 2^k d_G^{rk}\right)^{m},
\end{align}
where the summation with respect to $\{q_0,q_1,\ldots,q_l\}$ ($q_0\ge1, q_1\ge 1,\ldots, q_l\ge 1$) is equal to the $(m-l-1)$-multicombination from a set of $l+1$ elements: 
\begin{align} 
\sum_{\substack{q_0+q_1+\cdots+q_l=m\\q_0\ge1, q_1\ge 1,\ldots, q_l\ge 1}}=\multiset{l+1}{m-l-1} = \binom{m-1}{l} \label{comb_q_1_q_2__q_l}.
\end{align}
By applying the above upper bound to the inequality~\eqref{start_upp_counting}, we obtain the main inequality~\eqref{ineq:cluster_counting}.
This completes the proof. $\square$

 {~}

\hrulefill{\bf [ End of Proof of Proposition~\ref{prop:cluster_counting} ] }

{~}

\section{Proof of Theorem~\ref{main_theorem_QAMC_mutual_information_long}}

We here show the proof of Theorem~\ref{main_theorem_QAMC_mutual_information_long} which upper bounds the conditional mutual information in long-range interacting systems.
We rewrite the Hamiltonian with the power-law decay interaction by using the notations~\eqref{def:E_x_set} and \eqref{eq:ham_graph_sup2}:
\begin{align}
H=\sum_{\ed\in E_\infty} h_\ed =\sum_{l=1}^\infty \sum_{\ed\in E^{(l)}}  h_\ed  .
\end{align}
We here define $\tilde{g}_l$ as 
\begin{align}
\tilde{g}_l:=  \max_{v\in V}  \sum_{\ed\in E^{(l)}|\ed \ni v} \| h_\ed \|  .
\label{definition_tilde_g_l}
\end{align}
Then, the assumption~\eqref{long_range_interaction} implies 
\begin{align}
\sum_{l\ge R}^\infty \sum_{\ed\in E^{(l)} | \ed \ni v} \| h_\ed \| \le \sum_{l\ge R} \tilde{g}_l \le R^{-\alpha}.
\label{long_extensive_sum_g} 
\end{align}

We again show the statement that we would like to prove:
\begin{theorem} \label{main_theorem_QAMC_mutual_information_long_supp}
Let $A$, $B$ and $C$ be arbitrary subsystems in $V$ ($A,B,C \subset V$).
Then, under the assumption that the inverse temperature satisfies 
\begin{align}
\beta < \beta_c/11 = \frac{1}{88e^3 k},
\end{align}
the Gibbs state $\rho$ satisfies the approximate Markov property as follows:
\begin{align}
\mi_\rho(A:C|B)  \le \beta \min(|A|, |C|) \frac{11e^{1/k}/\beta_c}{1-11\beta/\beta_c} \dist_{A,C}^{-\alpha}   \label{eq:quantum_approximate_markov_chain_long_supp},
\end{align}
where we assume that $\dist_{A,C} \ge 2\alpha$.
\end{theorem}

\subsection{Details of the proof}

We start from Eq.~\eqref{Cluster_decomp_mutual}. By parametrizing the Hamiltonian as 
\begin{align}
H_{\vec{a}}=\sum_{\ed\in E_\infty} a_\ed h_\ed =\sum_{l=1}^\infty \sum_{\ed\in E^{(l)}} a_\ed h_\ed  ,
\end{align}
we have 
\begin{align}
\tilde{H}_{\vec{1}}(A:C|B)  
=& \sum_{m=1}^{\infty}\frac{1}{m!} \sum_{\ed_1,\ed_2,\ldots,\ed_m\in E_\infty} \prod_{j=1}^m \frac{\partial}{\partial a_{\ed_j}}\log(\tilde{\rho}_{\vec{a}}^{L}) \Bigl|_{\vec{a}=\vec{0}} \notag \\
=& \sum_{m=1}^{\infty}\frac{1}{m!} \sum_{l_1,l_2,\ldots,l_m=1}^{\infty} \sum_{\ed_1\in E^{(l_1)},\ed_2\in E^{(l_2)},\ldots, \ed_m\in E^{(l_m)}}  \prod_{j=1}^m \frac{\partial}{\partial a_{\ed_j}}\log(\tilde{\rho}_{\vec{a}}^{L}) \Bigl|_{\vec{a}=\vec{0}} \notag \\
=& \sum_{m=1}^{\infty}\frac{1}{m!} \sum_{l_0=m}^{\infty} \sum_{w \in \Cl_m(l_0)} n_w \Der_{w}  \tilde{H}_{\vec{a}}(A:C|B)\Bigl|_{\vec{a}=\vec{0}}, \label{Cluster_decomp_mutual_long}
\end{align}
where we define $ \Cl_m(l_0)\subset  \Cl_{\infty,m}$ as 
\begin{align}
\Cl_m(l_0) = \left \{w=\{\ed_1,\ed_2,\ldots,\ed_m\} \in\Cl_{\infty,m} \Biggl| \ed_j\in E^{(l_j)} , \ j=1,2,\ldots, m \quad  {\rm s.t. } \quad \sum_{j=1}^m l_j =l_0\right\}.
\end{align}
See Eq.~\eqref{def:E_x_set} and Sec.~\ref{subsub_sec:Preliminaries} for the definitions of $\Cl_{\infty,m}$ and $E^{(l)}$.

Next, from Eq.~\eqref{Cluster_decomp_mutual_long}, we can derive a similar statement to the proposition~\ref{prop: connceted_cluster_mutual}: 
\begin{align}
\tilde{H}_{\vec{1}}(A:C|B)  &=\sum_{m=1}^{\infty}\frac{1}{m!} \sum_{l_0=m}^{\infty} \sum_{w \in \Cl_m(l_0)} n_w \Der_{w}  \tilde{H}_{\vec{a}}(A:C|B)\Bigl|_{\vec{a}=\vec{0}} \notag \\
&= \sum_{m=1}^{\infty}\frac{1}{m!} \sum_{l_0 \ge \dist_{A,C}}   \sum_{w \in \Gc^{A,C}_{m}(l_0)} n_w \Der_{w}  \tilde{H}_{\vec{a}}(A:C|B)\Bigl|_{\vec{a}=\vec{0}} ,
\end{align}
where we define $\Gc^{A,C}_m(l_0)\subset \Gc^{A,C}_{\infty, m}$ as 
\begin{align}
\Gc^{A,C}_m(l_0) = \left \{w=\{\ed_1,\ed_2,\ldots,\ed_m\} \in\Gc^{A,C}_{\infty, m}\Biggl| \ed_j\in E^{(l_j)} , \ j=1,2,\ldots, m \quad  {\rm s.t. } \quad \sum_{j=1}^m l_j =l_0\right\}.
\end{align} 
Notice that we have $w \notin \Gc^{A,C}_{m}(l_0)$ if $ l_0 < \dist_{A,C}$ from the above definition. 

By following the same discussions in the derivation of Ineq.~\eqref{mutual_sum_decomp_log}, we obtain
\begin{align}
\| \tilde{H}_{\vec{1}}(A:C|B) \|
\le \sum_{v \in A} \sum_{m=1}^{\infty} \sum_{l_0 \ge \dist_{A,C}} 2 (4\beta)^m    \sum_{w \in \Gc^{v}_{r,m}(l_0)} \frac{n_w}{m!} \prod_{s=1}^{m}  N_{\ed_{s} | w} \|h_{\ed_s}\|     ,\label{mutual_sum_decomp_log2}
 \end{align}
where in this case, the summation of $v \in \partial A_r$ is replaced by $v \in A$ due to  $\partial A_\infty = A$ (see Eq.~\eqref{def_surface_region_supp}). 
Then, by using the inequality~\eqref{Prop:cluster_summation_region_L1}, obtain
\begin{align}
\sum_{w \in \Gc^{v}_{r,m}(l_0)} \frac{n_w}{m!} \prod_{s=1}^{m}  N_{\ed_{s} | w} \|h_{\ed_s}\| 
\le \frac{e^{1/k} (2e^{3} k)^m}{2} \sum_{l_1+l_2+\ldots+l_m=l_0} \prod_{j=1}^m \tilde{g}_{l_j} , \label{Prop:cluster_summation_region_L_long}
\end{align}
where we defined $\tilde{g}_{l}$ in Eq.~\eqref{definition_tilde_g_l}.
By combining the inequalities~\eqref{mutual_sum_decomp_log2} and \eqref{Prop:cluster_summation_region_L_long}, we obtain
\begin{align}
\| \tilde{H}_{\vec{1}}(A:C|B) \|
&\le \sum_{v \in A}\sum_{m=1}^{\infty}  \sum_{l_1+l_2 + \cdots + l_m \ge \dist_{A,C}}   e^{1/k}  (8 e^{3} k\beta)^m \prod_{j=1}^m \tilde{g}_{l_j}   . 
 \label{sum_long_extensive_power_law0}
 \end{align}
We can prove the following inequality (see Sec.~\ref{proof:sum_long_extensive_power_law} for the proof):
\begin{align}
 \sum_{l_1+l_2 + \cdots + l_m \ge l_0}\prod_{j=1}^m \tilde{g}_{l_j}  
 \le  11^{m} l_0^{-\alpha} 
 \label{sum_long_extensive_power_law}
 \end{align}
for arbitrary $l_0 \ge 2\alpha$.
By using the above inequality, we obtain 
\begin{align}
\sum_{m=1}^{\infty}  \sum_{l_1+l_2 + \cdots + l_m \ge \dist_{A,C}}e^{1/k}  (8 e^{3} k\beta)^m  \prod_{j=1}^m \tilde{g}_{l_j} 
\le  \dist_{A,C}^{-\alpha} \sum_{m=1}^{\infty} e^{1/k}  (11\beta/\beta_c)^m 
\le   \frac{11e^{1/k} \beta/\beta_c}{1-11\beta/\beta_c}  \dist_{A,C}^{-\alpha}.
\label{sum_long_extensive_power_law2}
 \end{align}
 By combining the inequalities~\eqref{sum_long_extensive_power_law0} and \eqref{sum_long_extensive_power_law2}, 
 we finally obtain 
\begin{align}
\| \tilde{H}_{\vec{1}}(A:C|B) \| \le \beta |A|  \frac{11e^{1/k}/\beta_c}{1-11\beta/\beta_c} \dist_{A,C}^{-\alpha}.\label{sum_long_extensive_power_law4}
 \end{align}
In the same way, we can derive the inequality such that $|A|$ is replaced by $|C|$ in \eqref{sum_long_extensive_power_law4}.
 By combining the above inequality with \eqref{cond_mutual_info_tilde_rho_upper_bound}, we prove Theorem~\ref{main_theorem_QAMC_mutual_information_long}.
$\square$

\subsubsection{Proof of the inequality~\eqref{sum_long_extensive_power_law}} \label{proof:sum_long_extensive_power_law}
For the proof, we start from the following form:
\begin{align}
\sum_{l_1+l_2 + \cdots + l_m  \ge l_0}  \prod_{j=1}^m\tilde{g}_{l_j} \le \eta_m l_0^{-\alpha}.
\label{start_form_tilde_g_l_j}
 \end{align}
We, in the following, construct a recurrence relation to determine $\eta_m$.
First, Eq.~\eqref{long_extensive_sum_g} immediately implies
\begin{align}
\sum_{l_1+l_2 + \cdots + l_m \ge l_0}  \prod_{j=1}^m \tilde{g}_{l_j}
 \le  \prod_{j=1}^m \sum_{l_j=1}^{\infty} \tilde{g}_{l_j} \le 1.
 \label{start_form_tilde_g_l_j_trivial_bound}
 \end{align}

Based on the inequalities~\eqref{start_form_tilde_g_l_j} and \eqref{start_form_tilde_g_l_j_trivial_bound}, we consider the case of $m+1$ as 
\begin{align}
\sum_{l_1+l_2 + \cdots + l_{m+1}  \ge l_0}  \prod_{j=1}^{m+1} \tilde{g}_{l_j}  
&\le \sum_{l_{m+1}=1}^\infty  \tilde{g}_{l_{m+1}} \sum_{l_1+l_2 + \cdots + l_m  \ge l_0- l_{m+1}}  \prod_{j=1}^{m}\tilde{g}_{l_j} \notag \\
&\le  \eta_m\sum_{l_{m+1}=1}^\infty  \tilde{g}_{l_{m+1}}\max\left[ ( l_0-l_{m+1})^{-\alpha} ,1\right] \notag \\
&\le  \eta_m\sum_{l=1}^{l_0-1}\tilde{g}_l   (l_0-l)^{-\alpha} + \eta_m \sum_{l\ge l_0} \tilde{g}_l \le  \eta_m\sum_{l=1}^{l_0-1}\tilde{g}_l   (l_0-l)^{-\alpha} + \eta_m l_0^{-\alpha},
\label{case_m+1_udner_assumption_gm}
 \end{align}
where the last inequality comes from the inequality~\eqref{long_extensive_sum_g} with $R=l_0$.
In order to upper-bound the first term, we decompose the summation as follows:
 \begin{align}
\sum_{l=1}^{l_0-1}\tilde{g}_l   (l_0-l)^{-\alpha} 
&=\left(  
\sum_{l\in [1,l_1 )} 
+\sum_{l\in [l_1 , l_2)}
+\sum_{l\in [l_2, l_3)}
+\sum_{l\in [l_3, l_0)}
 \right )\tilde{g}_l   (l_0-l)^{-\alpha} ,
\label{upper_bound_sum_tilde_g_l_l_0}
 \end{align} 
 for $\alpha >2$, where $l_1=\lceil l_0/\alpha \rceil$, $l_2=\lceil  l_0/2 \rceil$, $l_3=   \lceil l_0 -  l_0/\alpha \rceil$.
 For $\alpha \le 2$, we decompose as 
  \begin{align}
\sum_{l=1}^{l_0-1}\tilde{g}_l   (l_0-l)^{-\alpha} 
&=\left(  
\sum_{l\in [1,l_2 )} 
+\sum_{l\in [l_2, l_0)}
 \right )\tilde{g}_l   (l_0-l)^{-\alpha} .
\label{upper_bound_sum_tilde_g_l_l_0_2}
 \end{align} 
 
Next, for arbitrary choice of $[x,y)$ ($1\le x \le y \le l_0-1$), we have
 \begin{align}
\sum_{l\in [x,y)} \tilde{g}_l   (l_0-l)^{-\alpha}  \le  (l_0-y+1)^{-\alpha} \sum_{l\in [x,y)} \tilde{g}_l    \le (l_0-y+1)^{-\alpha} \sum_{l\ge x} \tilde{g}_l   
 \le (l_0-y+1)^{-\alpha} x^{-\alpha},
 \end{align} 
which reduces the inequality~\eqref{upper_bound_sum_tilde_g_l_l_0} to
 \begin{align}
\sum_{l=1}^{l_0-1}\tilde{g}_l   (l_0-l)^{-\alpha} 
\le& (l_0-\lceil  l_0/\alpha \rceil+1)^{-\alpha}  + (l_0-\lceil  l_0/2 \rceil+1)^{-\alpha} \lceil  l_0/\alpha \rceil^{-\alpha} \notag \\
&+ (l_0- \lceil l_0 -  l_0/\alpha \rceil +1)^{-\alpha} \lceil  l_0/2 \rceil^{-\alpha} + \lceil l_0 -  l_0/\alpha \rceil^{-\alpha} \notag \\
\le & 2(l_0-l_0/\alpha )^{-\alpha} + 2 (l_0/2)^{-\alpha} (l_0/\alpha)^{-\alpha}  \notag \\
=&2 l_0^{-\alpha} \left[ \frac{1}{(1-1/\alpha )^{\alpha}} + \left(\frac{2\alpha}{l_0}\right)^{\alpha}\right]
\le 10l_0^{-\alpha}
\label{l_1_l0_1_tilde_g_lo_alpha_ge2}
 \end{align} 
for $\alpha >2$, where we use $1/(1-1/x)^x \le 4$ for $x\ge 2$ and $l_0 \ge 2\alpha$ from the condition of the theorem.
For $\alpha \le 2$, we also obtain
 \begin{align}
\sum_{l=1}^{l_0-1}\tilde{g}_l   (l_0-l)^{-\alpha}  \le 2 (l_0/2)^{-\alpha} \le 8 l_0^\alpha
\label{l_1_l0_1_tilde_g_lo_alpha_le2}
 \end{align} 
 from the decomposition~\eqref{upper_bound_sum_tilde_g_l_l_0_2}, where we use $2^\alpha \le 4$ for $\alpha \le 2$.
 
By applying the inequalities~\eqref{l_1_l0_1_tilde_g_lo_alpha_ge2} and \eqref{l_1_l0_1_tilde_g_lo_alpha_le2} to the inequality~\eqref{case_m+1_udner_assumption_gm}, we obtain
\begin{align}
\sum_{l_1+l_2 + \cdots + l_{m+1}  \ge l_0}  \prod_{j=1}^{m+1} \tilde{g}_{l_j}  
&\le  11 \eta_m l_0^{-\alpha},
 \end{align}
which gives rise to
\begin{align}
\eta_{m+1}\le  11  \eta_m  . \label{ineq_eta_m+1_eta_m2}
 \end{align}
This yields the inequality~\eqref{sum_long_extensive_power_law}.
This completes the proof. $\square$


\end{widetext}

\end{document}